\documentclass[twocolumn,twocolappendix]{aastex63}

\usepackage[version=4]{mhchem}
\usepackage{amsmath,amssymb}
\received{}
\revised{}
\accepted{}
\submitjournal{ApJ}

\shorttitle{Different degrees of nitrogen and carbon depletion}
\shortauthors{Furuya et al.}
\newcommand{\fdep}[1]{f_{\rm depl}({\rm #1})}
\newcommand{\cogas}{{\rm [C/O]_{gas}}}

\begin{document}

\title{Different degrees of nitrogen and carbon depletion in the warm molecular layers of protoplanetary disks}

\correspondingauthor{Kenji Furuya}
\email{kenji.furuya@nao.ac.jp}

\author[0000-0002-2026-8157]{Kenji Furuya}
\affiliation{National Astronomical Observatory of Japan, Osawa 2-21-1, Mitaka, Tokyo 181-8588, Japan}
\author[0000-0002-0226-9295]{Seokho Lee}
\affiliation{Korea Astronomy and Space Science Institute, 776 Daedeok-daero, Yuseong-gu, Daejeon 34055, Republic of Korea}
\author[0000-0002-7058-7682]{Hideko Nomura}
\affiliation{National Astronomical Observatory of Japan, Osawa 2-21-1, Mitaka, Tokyo 181-8588, Japan}
\affiliation{Department of Astronomical Science, The Graduate University for Advanced Studies (SOKENDAI), Osawa, Mitaka, Tokyo 181-8588, Japan}
%\author{Yuri Aikawa}
%\affiliation{Department of Astronomy, The University of Tokyo, Bunkyo-ku, Tokyo 113-0033, Japan}

%% Note that the \and command from previous versions of AASTeX is now
%% depreciated in this version as it is no longer necessary. AASTeX 
%% automatically takes care of all commas and "and"s between authors names.

%% AASTeX 6.3 has the new \collaboration and \nocollaboration commands to
%% provide the collaboration status of a group of authors. These commands 
%% can be used either before or after the list of corresponding authors. The
%% argument for \collaboration is the collaboration identifier. Authors are
%% encouraged to surround collaboration identifiers with ()s. The 
%% \nocollaboration command takes no argument and exists to indicate that
%% the nearby authors are not part of surrounding collaborations.

%% Mark off the abstract in the ``abstract'' environment. 
\begin{abstract}
%We show that nitrogen can become the third most abundant element in the gas phase after hydrogen and helium in the warm molecular layers of protoplanetary disks.
Observations have revealed that the elemental abundances of carbon and oxygen in the warm molecular layers of some protoplanetary disks are depleted compared to those is the interstellar medium by a factor of $\sim$10-100.
Meanwhile, little is known about nitrogen.
To investigate the time evolution of nitrogen, carbon, and oxygen elemental abundances in disks, we develop a one-dimensional model that incorporates dust settling, turbulent diffusion of dust and ices, as well as gas-ice chemistry including the chemistry driven by stellar UV/X-rays and the galactic cosmic rays.
We find that gaseous CO in the warm molecular layer is converted to \ce{CO2} ice and locked up near the midplane via the combination of turbulent mixing (i.e., the vertical cold finger effect) and ice chemistry driven by stellar UV photons.
On the other hand, gaseous \ce{N2}, the main nitrogen reservoir in the warm molecular layer, is less processed by ice chemistry, and exists as it is.
Then the nitrogen depletion occurs solely by the vertical cold finger effect of \ce{N2}.
As the binding energy of \ce{N2} is lower than that of CO and \ce{CO2}, the degree of nitrogen depletion is smaller than that of carbon and oxygen depletion, leading to higher elemental abundance of nitrogen than that of carbon and oxygen.
This evolution occurs within 1 Myr and proceeds further, when the $\alpha$ parameter for the diffusion coefficient is $\sim$10$^{-3}$.
Consequently, the \ce{N2H+}/CO column density ratio increases with time.
%The degree of carbon and nitrogen depletion are independent of the cosmic-ray ionization rate at least for the 1 Myr time evolution. 
How the vertical transport affects the midplane ice composition is briefly discussed.
\end{abstract}

%% Keywords should appear after the \end{abstract} command. 
%% See the online documentation for the full list of available subject
%% keywords and the rules for their use.

%% From the front matter, we move on to the body of the paper.
%% Sections are demarcated by \section and \subsection, respectively.
%% Observe the use of the LaTeX \label
%% command after the \subsection to give a symbolic KEY to the
%% subsection for cross-referencing in a \ref command.
%% You can use LaTeX's \ref and \label commands to keep track of
%% cross-references to sections, equations, tables, and figures.
%% That way, if you change the order of any elements, LaTeX will
%% automatically renumber them.
%%
%% We recommend that authors also use the natbib \citep
%% and \citet commands to identify citations.  The citations are
%% tied to the reference list via symbolic KEYs. The KEY corresponds
%% to the KEY in the \bibitem in the reference list below. 

%% Appendix material should be preceded with a single \appendix command.
%% There should be a \section command for each appendix. Mark appendix
%% subsections with the same markup you use in the main body of the paper.

%% Each Appendix (indicated with \section) will be lettered A, B, C, etc.
%% The equation counter will reset when it encounters the \appendix
%% command and will number appendix equations (A1), (A2), etc. The
%% Figure and Table counter will not reset.

\section{Introduction}
%One of the most fundamental parameters of the planet formation is the gas mass of protoplanetary disks.
%In the interstellar medium (ISM), CO is the second most abundant molecule in the gas phase with an abundance of $\sim$10$^{-4}$ with respect to \ce{H2} \citep{frerking82}, 
%and is commonly used as a tracer of molecular gas mass.
%Constraining disk gas mass with CO observations is, however, not straightforward, due to the uncertainties in the CO/\ce{H2} abundance ratio in protoplanetary disks.
%elemental abundance,

%CO depletion problem. what is the gas mass tracer?

Formation of planets occurs in protoplanetary disks.
The elemental compositions of gas and solids in disks directly set the elemental compositions of forming planets in the disks.
Understanding the elemental compositions of gas and solids in disks is thus the first step in understanding how planets acquire their compositions \citep[e.g.,][]{oberg21}.
%planet compositionとformation location \citep[e.g.,][]{oberg11}.
%Not only the disk midplane, but also the disk atomosphere can be important, meridional flow (teague)
%it is important to understand the distribution of elements inside disks
%The distributions of molecules across disks regulate the elemental compositions of planets, including C/N/O/S ratios and metallicity (O/H and C/H), as well as access to water and prebiotically relevant organics.

Recent observations have found that elemental abundances of volatile carbon and oxygen in some disks 
are different from those in the interstellar medium (ISM).
In the three disks (TW Hya, DM Tau, and GM Aur), 
where gas mass is measured with the HD line observations \citep{bergin13,mcclure16}, the CO abundance in the warm molecular layer ($T \gtrsim$ 20 K) is lower than the canonical value of $10^{-4}$ by factors of 10--100 \citep[e.g.,][]{favre13,zhang19,zhang21}.
%, although the absolute value of the estimated CO abundance is model-dependent \citep{ruaud22}.
As CO is the major carrier of volatile carbon, a reasonable hypothesis is that volatile elemental carbon is depleted in the gas phase by similar factors.
Survey observations of disks in nearby star-forming regions have found that CO emission is much weaker than expected, even when the freeze-out of CO onto dust grains and photodissociation by stellar UV photons are considered in disk models \citep{ansdell16,long17,miotello17}.
The faint CO emission may indicate that the CO abundance in the warm molecular layer in disks is typically lower than the canonical value. Alternatively, it may indicate that gas mass of \ce{H2} is lower than expected in the disks.
The timescale of carbon (and oxygen) depletion has been suggested to be fast ($\sim$1 Myr), 
based on the observations of CO isotopologues towards disks in different evolutionary stages/ages \citep{bergner20b,zhang20}.
In addition, some disks show brighter \ce{C2H} emission than expected, indicating the C/O ratio higher than unity at least in the \ce{C2H} emitting regions \citep{bergin16,miotello19,bosman21}.
The high C/O ratio indicates that the degree of oxygen depletion is more significant than that of carbon.

There are some proposed mechanisms to explain the depletion of volatile carbon and oxygen in the warm molecular layer: conversion of CO into less volatile species through chemistry driven by cosmic-rays/X-rays \citep[e.g.,][]{bergin14,furuya14}, so-called vertical cold finger effect \citep[e.g.,][]{meijerink09,kama16,xu17}, and a combination of both of these \citep{krijt20}.
The vertical cold finger effect is the phenomena as follows;
Turbulence brings the gaseous molecules from the warm molecular layer to the disk midplane, 
where the dust temperature is low and molecules are frozen out onto large dust grains.
As large dust grains are not fully coupled with the turbulent gas motion, they are settled close to the midplane, trapping ices on them.
As a result, the abundances of gaseous molecules in the disk atmosphere can decrease with time.

% inner disk is not necessarily carbon-rich, dust trap? chemical conversion?

Compared to carbon and oxygen, our understanding of the nitrogen elemental abundance in disks is rather limited.
The main carrier of nitrogen in the warm molecular layer is expected to be \ce{N2} or atomic N in the gas phase \citep[e.g.,][]{li13}, which are not directly observable.
Some observational studies have concluded that the degree of nitrogen depletion is significantly smaller than that of carbon and oxygen. 
\citet{cleeves18} concluded that carbon and oxygen are depleted in the observable gas of the IM Lup disk compared to the ISM elemental abundances by a factor of $\sim$20 or even larger, 
whereas nitrogen is not significantly depleted (up to a factor of four compared to the ISM elemental abundance), based on the observations of CO, \ce{C2H}, and HCN and the physical-chemical modeling of the disk.
The estimated age of IM Lup is 0.5--1 Myr \citep{mawet12}, 
again indicating the timescale of carbon and oxygen depletion is fast.
\citet{anderson19} found that the \ce{N2H+}/CO flux ratio in 5–11 Myr old disks in Upper Scorpius is higher 
than that in $\lesssim$2 Myr old disks, including the IM Lup disk.
The high \ce{N2H+}/CO flux ratio in the older disks may indicate  that \ce{N2}, which is the parent molecule of \ce{N2H+}, is less impacted by the physical and chemical processes in disks, and remains in the gas phase for longer timescale than CO.
It also may indicate that the degree of carbon depletion continues to be enhanced over time after $\sim$2 Myr.

In addition to the elemental composition in disks, 
nitrogen abundance is relevant to the measurements of the disk gas mass;
\citet{trapman22} showed that the combination of \ce{C^18O} line and \ce{N2H+} line can be a good prove of disk gas mass, assuming that \ce{N2} is not depleted in the disk gas.
One of the biggest uncertainties to measure the disk gas mass is the CO abundance in the warm molecular layer, as it is not necessarily the canonical value of 10$^{-4}$ as mentioned above. 
As \ce{N2H+} is destroyed by CO, the \ce{C^18O}/\ce{N2H+} flux ratio is useful to constrain the CO abundance in the warm molecular layer \citep{trapman22}, if \ce{N2} is not depleted significantly in contrast to CO.

Theoretically, the elemental abundance of nitrogen in the warm molecular layer can be reduced as well as that of carbon.
CO and \ce{N2} have similar volatility; laboratory experiments showed that the binding energy of \ce{N2} is slightly lower than that of CO, and their sublimation temperatures should be within a few degrees of each another \citep{fayolle16}.
Then, the \ce{N2} abundance in the warm molecular layer can be reduced by the vertical cold finger effect like CO.
In addition, the chemical conversion of \ce{N2} to less volatile species (\ce{NH3}) driven by cosmic-rays/X-rays can occur after significant CO depletion \citep[when the CO abundance becomes lower than $\sim$10$^{-5}$;][]{furuya14}.

For quantitative understanding of the evolution of C/N/O elemental abundances in disks,
models, that consider both gas-ice chemistry and the vertical cold finger effect are required.
Several previous studies have investigated detailed gas-ice chemistry in turbulent disks \citep[e.g.,][]{semenov11,furuya13,furuya14}.
In these studies, however, dust grains were assumed to be dynamically well coupled with gas,
and therefore the vertical cold finger effect was not included.
Other studies focused only on carbon and oxygen, and chemical processes other than adsorption and desorption of molecules were not considered \citep{xu17,krijt18}.
\citet{krijt20} combined the dust evolution and grain surface chemistry of carbon and oxygen driven by cosmic rays, but ignored the chemistry driven by stellar UV photons; 
as growth and settling of dust grains proceed, UV photons penetrate deeper into the disks, driving chemistry induced by UV photons \citep[e.g.,][]{akimkin13,vanclepper22}.

\begin{figure*}[ht]
\plotone{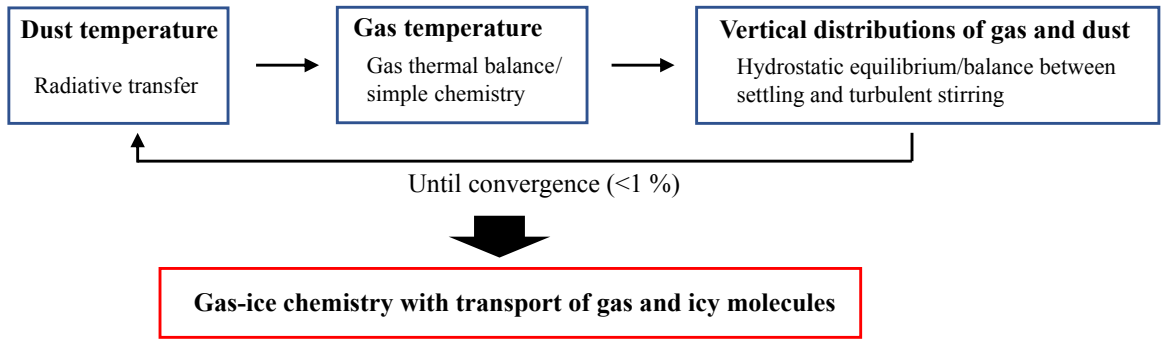}
\caption{An outline of the modeling processes.
}
\label{fig:flowchart}
\end{figure*}

In this study, we investigate the time evolution of carbon-, nitrogen-, and oxygen-bearing species in a disk, considering the vertical cold finger effect and gas-ice chemistry, which includes the chemistry driven by stellar UV/X-rays and the galactic cosmic rays.
%We show that considering CR-driven chemistry and stellar UV chemistry are important.
%We show that the degree of nitrogen depletion is less significant compared to those of carbon and oxygen, and nitrogen in the warm molecular layer can be more abundant than carbon and oxygen within 1 Myr depending on the strength of the turbulence.
%The key difference between carbon and nitrogen is the different reactivities of CO and \ce{N2} against the chemistry driven by stellar UV photons.
The rest of this paper is organized as follows.
Our numerical model is described in Section \ref{sec:model}, and the numerical results are presented in Section \ref{sec:result}.
We discuss the timescale of elemental depletion, the temporal evolution of column densities of gas-phase species, and the midplane ice composition in Section \ref{sec:discuss}.
Our findings are summarized in Section \ref{sec:summary}.

\section{Model} \label{sec:model}
\subsection{Numerical setup}
Figure \ref{fig:flowchart} shows an outline of our modeling processes.
We first obtain the self-consistent solution of the density and temperature profiles of the gas and dust in the vertical direction of a disk, irradiated by stellar UV and X-rays.
The gas surface density profile in the disk is taken from the TW Hya disk model 
developed in \citet{cleeves15} with the global dust-to-gas mass ratio of 0.01.
The size distribution of dust in an entire disk 
follows a power law with an index of -3.5 with minimum ($a_{\rm min}$) and maximum radii ($a_{\rm max}$) of 0.1 $\mu$m and 1 mm, respectively.
The dust size distribution is discretized into 50 logarithmically spaced bins.

Dust temperature is obtained from the radiative equilibrium using RADMC-3D code \citep{dullemond12}.
Gas temperature is obtained by solving the steady-state balance between heating and cooling, considering simplified chemistry \citep{lee21}.
Gas density distribution is determined from the hydrostatic equilibrium, 
while the distribution of dust with different radius is obtained by solving the steady-state balance between settling and turbulent stirring \citep[e.g.,][]{takeuchi02}.
Our vertical gas distribution is different from that in \citet{cleeves15}, which was calculated assuming a Gaussian profile and the parametric gas scale height.
We have confirmed that the degree of carbon and nitrogen depletion discussed in Section \ref{sec:result} does not change significantly whether using the vertical gas distribution from the hydrostatic equilibrium or the parametric distribution in \citet{cleeves15}.
The dust settling velocity and the diffusion coefficient for turbulent stirring are given later in this section.

The input stellar spectrum is composed of the observed UV spectrum of TW Hya \citep{dionatos19}, added to a Black body component with the effective temperature of 4110 K \citep{andrews12}.
As our model is one-dimensional plane-parallel, the stellar flux at a radius $R$ from the star is assumed to be
absorbed at the surface of the disk and half of the flux diffuses into the vertical direction towards the midplane.
The stellar flux at wavelength $\nu$ at the surface of the disk ($z_{\rm s}(R) = 0.8R$ in our models) is given by
\begin{align}
F(\nu, R) = \frac{1}{2}\frac{L_*(\nu)}{R^2 + z_{\rm s}(R)^2},
\end{align}
where $L_*(\nu)$ is the stellar spectrum.
The stellar flux at each height $z$ is calculated by RADMC-3D \citep{dullemond12}.
The dust opacity for UV and longer wavelengths is calculated with \texttt{dsharp\_opac} package from \citet{birnstiel18}, assuming the dust composition listed in their Table 1.
In addition to dust, we consider polycyclic aromatic hydrocarbons (PAHs) as a source of UV opacity \citep{weingartner01} and gas heating.
We assume that a PAH radius of $\sim$5 \AA \citep{weingartner01} and PAHs are always dynamically coupled with gas.
The average Galactic PAH-to-dust mass ratio is 4.6 \% \citep{draine07}, 
while the PAH abundance in disks is lower than that value by a factor of 10-100 \citep{geers06}.
In our models, PAH-to-dust mass ratio in an entire disk is assumed to be 0.05 \%, i.e., 100 times lower than the Galactic average value. 
X-ray ionization rate is calculated by the analytical formula \citep{igea99,bai09}, with the stellar X-ray luminosity of 10$^{30}$ erg s$^{-1}$.
The non-attenuated cosmic-ray (CR) ionization rate is set to be 10$^{-18}$ s$^{-1}$ and the CR ionization rate of each disk position is calculated with the attenuation column of 96 g cm$^{-2}$ \citep{umebayashi81}.

Next, we solve gas and ice chemistry, considering the vertical transport of gas and ice, under the physical conditions obtained in the first step:
%\begin{linenomath*}
\begin{align}
\frac{\partial n_i}{\partial t} + \frac{\partial( n_i V_T)}{\partial z} - \frac{\partial \phi_i}{\partial z} = P_i - L_i, \label{eq:basic}\\
\phi_i = -n_{\rm H} \frac{D_{z}}{S_c} \frac{\partial}{\partial z}\left(\frac{n_i}{n_{\rm H}}\right),  \label{eq:phi}
\end{align}
%\end{linenomath*}
where $n_i$ is the number density of species $i$, and $z$ is the height from the midplane.
$P_i$ and $L_i$ represent the production rate and the loss rate, respectively, of species $i$ by chemical reactions (see Section \ref{sec:chem}).
The second term in the left-hand side of Equation (\ref{eq:basic}) describes the vertical settling of dust, while the third term describes the turbulent diffusion.
The dust setting velocity is $V_T = -\Omega_K \tau_s z$, 
where $\Omega_K$ is the Keplerian orbital frequency \citep{nakagawa81}.
%Assuming Epstein regime, the dimensionless stopping time $\tau_s$ is defined as $\tau_s = \rho_{\rm int}\Omega_K a/\rho_g c_s$,
Assuming Epstein regime, the dimensionless stopping time $\tau_s$ is defined as $\tau_s = \rho_{\rm int}\Omega_K a/\rho_g c_s$,
where $\rho_{\rm int}$ is the internal density of dust (1.68 g cm$^{-3}$), $a$ is the radius of dust, $\rho_g$ is the mass density of gas, and $c_s$ is the local sound velocity.
The diffusion coefficient is parameterized as $D_z =  \alpha c_s^2/\Omega_K$ \citep{shakura73}, where $\alpha$ is assumed to be 10$^{-3}$ in this study, unless otherwise stated. 
The Schmidt number, $S_c$, describes the strength of coupling between gas and dust and is given by $1 + \tau_s^2$ \citep{youdin07}.
For gas phase species, $V_T = 0$ and $S_c = 1$. 
%first solve equation without chemistry 

%Initially, we assu
%Initial elemental abundances are taken from \citet{wakelam08}, where the C/N and C/O elemental abundance ratios are $\sim$2 and $\sim$0.5, respectively.
%In order to obtain the initial molecular abundance for the disk, we simulate 
%molecular evolution for 1 Myr under prestellar core conditions;
%the gas density, temperature, and the visual extinction are
%set to $2\times10^4$ cm$^{-3}$, 10 K, and 10 mag, respectively.

For the numerical integration of Equation \ref{eq:basic}, 
the first order operator splitting method is employed; 
we first integrate the equation between times $t$ and $t+\Delta t$ without the source terms of  chemical reactions.
Next, we integrate the equation without the transport terms (i.e., only with the terms of chemical reactions) between times $t$ and $t + \Delta t$ at each height $z$.
In the time integration of the chemical reaction terms, we re-bin the dust size distribution into 8 bins (that is, icy species on 8 different dust populations are distinguished from each other) to shorten the computational time.
The dust sizes for each of the 8 dust populations are ($a_{\rm min, \,j}$, $a_{\rm max, \,j}$) =  
(0.10 $\mu$m, 0.32 $\mu$m), (0.32 $\mu$m, 1.0 $\mu$m), (1.0 $\mu$m, 3.2 $\mu$m), (3.2 $\mu$m, 10 $\mu$m), (10 $\mu$m, 32 $\mu$m), (32 $\mu$m, 100 $\mu$m), (100 $\mu$m, 320 $\mu$m), and (320 $\mu$m, 1 mm), where $a_{\rm max, \,j}$ and $a_{\rm min, \,j}$ are the maximum and minimum radius for each dust bin, respectively.
Even if we used the larger number (e.g., 12) of dust bins for re-binning, numerical results presented in Section \ref{sec:result} remained almost unchanged.

The averaged radius ($\overline{a_j}$) of each dust bin for calculations of the chemistry are defined as follows \citep{vasyunin11}:
%\begin{linenomath*}
\begin{equation}
\overline{a_j}(z) = \frac{\int^{a_{\rm max, \,j}}_{a_{\rm min, \,j}} a^3 f(a, z)da}{\int^{a_{\rm max, \,j}}_{a_{\rm min, \,j}} a^2 f(a, z)da},
\end{equation}
%\end{linenomath*}
where $f$ is the number density of grains with size between $a$ and $a + da$.
Dividing the total mass of each dust population per unit volume ($\rho_d(a_j)$) 
by the mean mass of one dust grain ($4 \pi \overline{a_j}^3 \rho_{\rm int}/3$), 
we can obtain the number of grains with the averaged radius of $\overline{a_j}$ per unit volume.
The temperature for each dust bin ($\overline{T_d}$) is defined by the area-weighted average temperature \citep{gavino21}, 
because the rate coefficients of gas-dust interactions are proportional to the surface area in general:
%\begin{linenomath*}
\begin{equation}
\overline{T_d}(\overline{a_j}, z) = \frac{\int^{a_{\rm max, \,j}}_{a_{\rm min, \,j}} T_d(a, z)a^2 f(a, z)da}{\int^{a_{\rm max, \,j}}_{a_{\rm min, \,j}} a^2 f(a, z)da}.
\end{equation}
%\end{linenomath*}

\subsection{Chemical reaction network} \label{sec:chem}
The chemical network used in this work is based on that in \citet{furuya14},
which includes gas-phase reactions, interaction between gas and (icy) grain surfaces, and grain surface reactions.
UV photodissociation/photoionization rates are calculated by convolving the local radiation field and the photodissociation/photoionization cross sections \citep{heays17}.
The self-shielding and mutual shielding factors for the photodissociation of \ce{H2} and CO are taken from \citet{visser09}.
Shielding factors for \ce{N2} photodissociation are taken from \citet{li13}.
%X-ray chemistry is calculated in the same way as in \citet{furuya13}.
The binding energies of CO, \ce{CO2}, \ce{H2O}, and \ce{N2} are set at 1150 K, 2600 K, 5700 K, and 1000 K, respectively.

\subsection{Initial compositions} \label{sec:init}
Initially, we assume spatially uniform volatile elemental abundances throughout the disk.
Our underlying initial elemental abundances for H:He:C:N:O:Na:Mg:Si:S:Fe
are 1.00:9.75(-2):1.00(-4):6.00(-5):2.20(-4):2.25(-9):1.09(-8):9.74(-9):9.14(-8):2.74(-9), respectively, where $a(-b)$ means $a\times10^{-b}$.
Then the initial elemental [C/N] and [C/O] ratios are $\sim$2 and $\sim$0.5, respectively.
For initial molecular abundances, we assume that 
all hydrogen is in \ce{H2},
all carbon is in CO,
the remaining oxygen is in \ce{H2O},
and all nitrogen is in \ce{N2}.
The other elements are assumed to exist as either atoms or atomic ions in the gas phase.
All CO and \ce{N2} are assumed to be initially present in the gas phase.
We determine the initial partitioning of \ce{H2O} among ice mantles of 8 different dust populations as follows.
All \ce{H2O} is assumed to be present in ice mantles, and the number density of \ce{H2O} ice on each dust population are assumed to scale with $\rho_d(\overline{a_j}, z)/\Sigma_k \rho_d(\overline{a_k},z)$, i.e., larger sized dust grains have the larger amount of \ce{H2O} ice.
%This assumption would be verified as larger dust grains are formed by the coagulation of smaller dust grains.
%Then we compare the adsorption timescale with the thermal desorption timescale for \ce{H2O} for each dust population.
%If the former is smaller than the latter for some dust populations, we assume that these dust populations do not have \ce{H2O} in ice mantles, and \ce{H2O} on those dust populations are moved to the gas phase.

Our two-step approach assumes that large dust grains are already settled and the depletion of elemental carbon, nitrogen, and oxygen in the disk upper layers have not occurred yet at the beginning of our simulations.
The first assumption would be verified, because observations of Class 0/I sources have found that
their inner ($<$1000 au) dense regions already contain mm-sized dust grains \citep[e.g.,][]{miotello14,harsono18,galametz19}.
In addition, the presence of dust ring structures observed in some Class 0/I disks may indicate that dust coagulation and settling have already started in the young disks \citep[e.g.,][]{sheehan18,nakatani20,ohashi21}.
The second assumption would be justified as well, at least for carbon and nitrogen;
Class 0/I disks are warmer than Class II disks (i.e., the vertical cold finger effect is less efficient) with no signs of CO freeze-out in the inner $\sim$100 au \citep{vanthoff20} and with  the CO abundance close to the canonical value \citep{zhang20}.
As \ce{N2} is more volatile than CO, it would be reasonable to assume \ce{N2} is entirely present in the gas phase.

The situation for water (i.e., oxygen) is less clear.
The H$_2^{18}$O emission has been detected in the inner warm ($\gtrsim$100 K) regions (inner envelope and/or a disk), where water ice has sublimated, of several Class 0 sources at spatial scales of 100 au \citep{persson12,jensen19}.
\citet{harsono20} reported non-detection of H$_2^{18}$O emission toward Class I disks at scales of 100 au, 
possibly because abundant large (mm-sized or even larger) dust grains suppress the emission from inner warm disk regions.
Theoretically speaking, one can consider two extreme scenarios about water in Class 0/I disks.
In one scenario, due to thermal desorption and/or photodesorption of water ice, followed by photodissociation of water in the gas phase, 
water ice on dust grains are uncovered in the upper layers of Class 0/I disks (still, water ice can exist close to the midplane).
As a result, dust coagulation and settling would not cause the depletion of volatile oxygen in the disk upper layers.
In this case, the volatile oxygen abundance would be uniformly $\sim$10$^{-4}$ in Class 0/I disks (oxygen in CO is not counted here).
In another scenario, both thermal desorption and photodesorption are negligible, and coagulation and settling of dust covered by water ice cause the depletion of volatile oxygen in the disk upper layers.
In this case, the volatile oxygen abundance scales with the local dust-to-gas mass ratio in Class 0/I disks.
Reality would be somewhere between the two extremes, depending on stellar properties and disk physical structures.
The former scenario is assumed in our fiducial models, while the latter scenario is discussed in Section \ref{sec:compare}.

\begin{figure*}[ht]
%    \centering
    \plotone{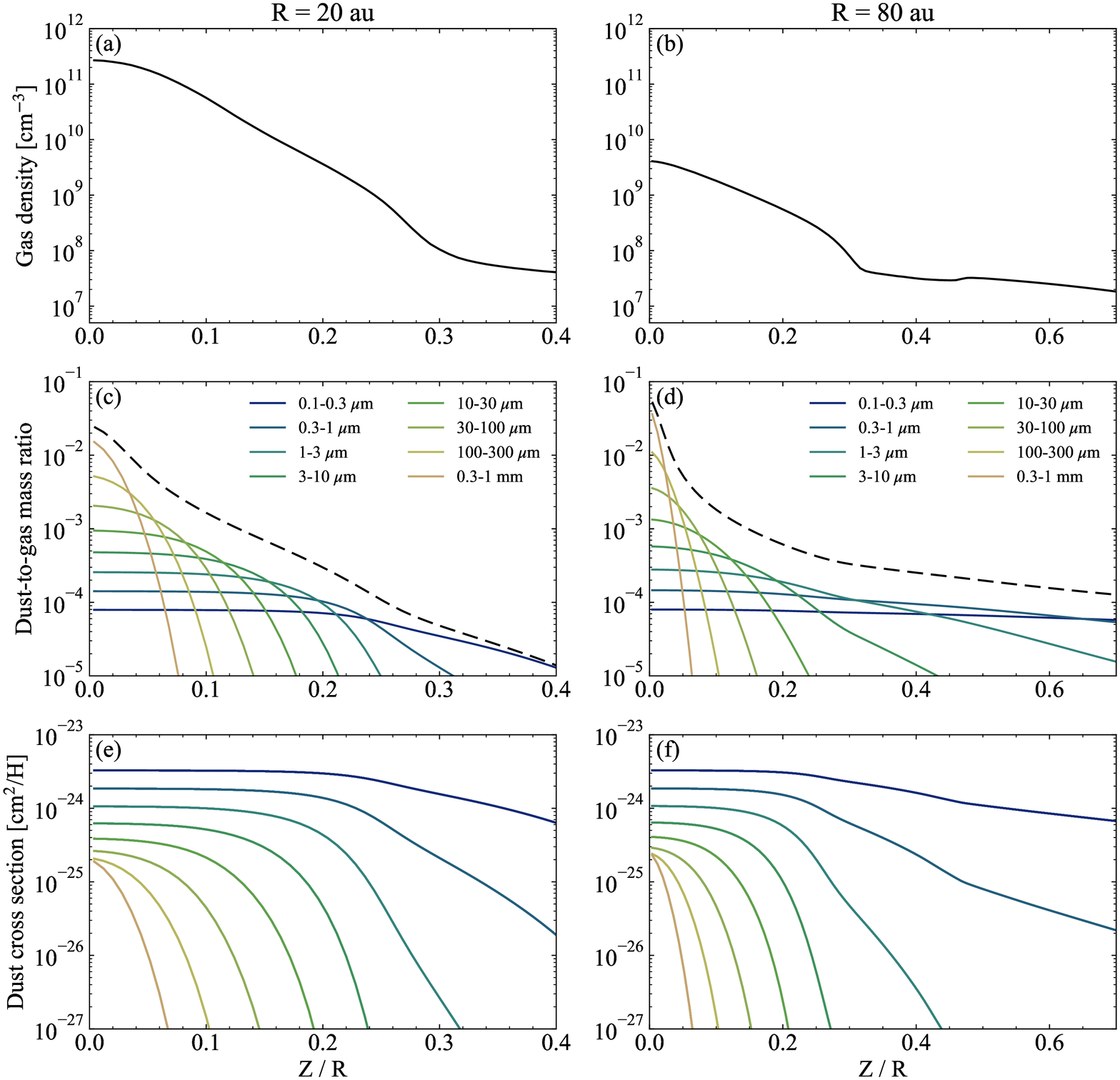}
    \caption{Vertical physical structures of a disk at $R = 20$ au (left panels) and 80 au (right panels) as functions of $Z/R$. (a, b) Gas density. (c, d) Dust-to-gas mass ratio for the 8 dust populations. {\rm Black dashed lines represent the sum of the dust-to-gas mass ratio for the 8 dust populations.} (e,f) Total cross section for the 8 dust populations per hydrogen nuclei. For comparison, when the size of all dust grains is 0.1 $\mu$m and the dust-to-gas mass ratio is 0.01, the total cross section is $\sim 3 \times 10^{-22}$ cm$^{-2}$ per hydrogen nuclei.}
    \label{fig:phys}
\end{figure*}

\begin{figure*}[ht]
%    \centering
    \plotone{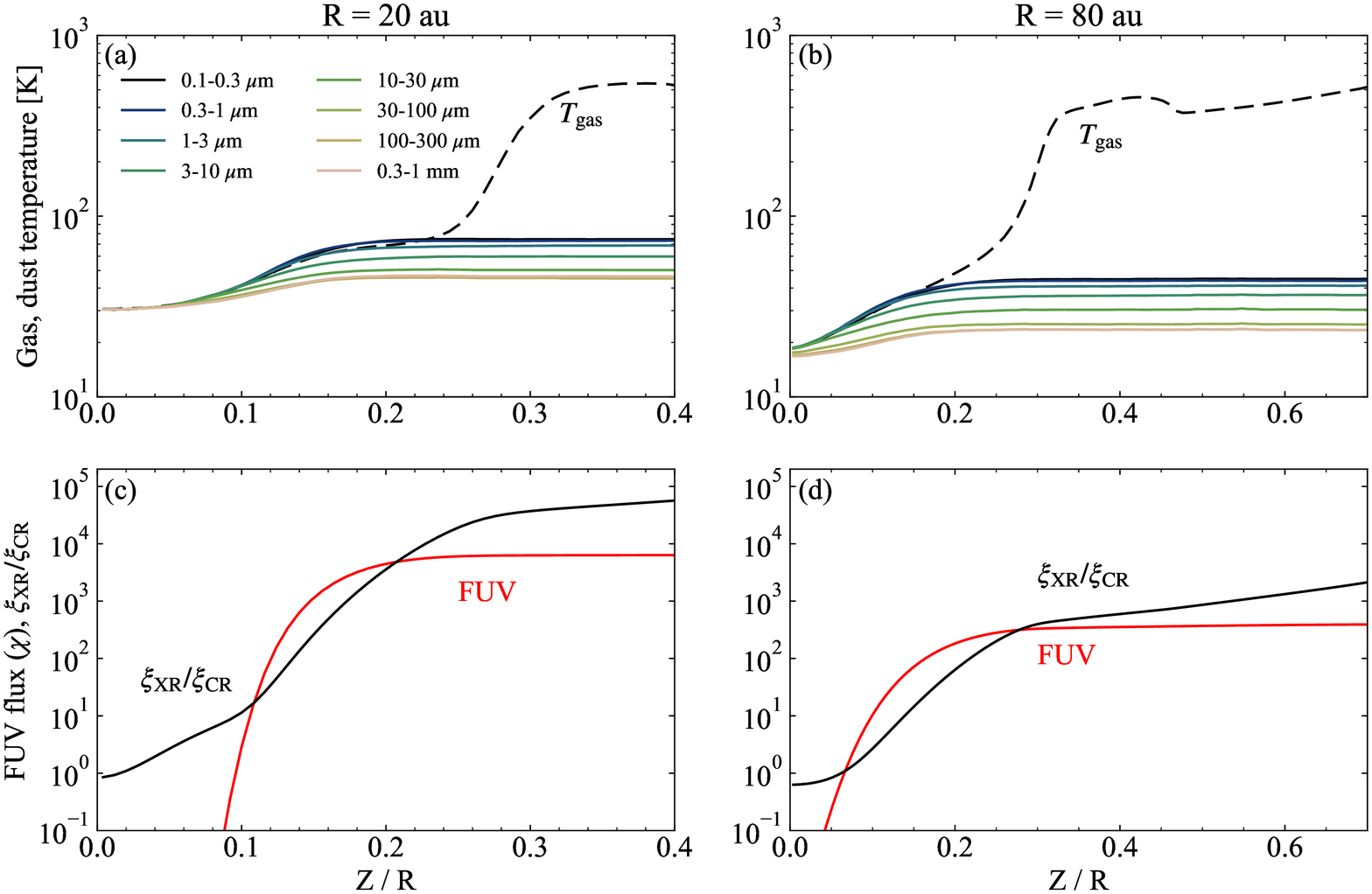}
    \caption{Similar to Figure \ref{fig:phys}, but for temperatures and UV/X-ray radiation fields.
    (a, b) Gas temperature (dashed black line) and dust temperature for the 8 dust populations (solid lines). (c, d) Wavelength-integrated FUV flux normalized by the Draine field \citep[$1.6 \times 10^{-3}$ erg cm$^{-2}$ s$^{-1}$;][]{draine78} (red) and the X-ray ionization rate normalized by the non-attenuated cosmic-ray ionization rate, which is set to 10$^{-18}$ s$^{-1}$ (black).}
    \label{fig:phys2}
\end{figure*}

\section{Results} \label{sec:result}
\subsection{Vertical distributions of molecular abundances} \label{sec:result1}

Figures \ref{fig:phys} and \ref{fig:phys2} show the vertical physical structure of the disk at $R = 20$ au (inside the CO snowline) and $R = 80$ au (outside the CO snowline) as functions of $z/R$.
Lager dust grains tend to settle closer to the midplane \citep[e.g.,][]{dullemond04}, 
and the dust mass is dominated by population with the largest size ($\lesssim$1 mm).
However, even at the disk midplane, the total cross section of dust, which is relevant with the rates of gas-dust interaction, is dominated by the population with the smallest size ($<$1 $\mu$m).
The dust temperature of larger dust grains is colder than that of smaller ones in the disk upper layers ($\gtrsim$0.1 $z/R$), reflecting the different wavelength-dependency of the opacity of different sized dust grains \citep[e.g.,][]{wolf03}.
Near the disk midplane, where even the sub-mm radiation becomes optically thick, the dust temperature is independent of its size.
As discussed in the Appendix \ref{app:vertical_cold_finger}, the size-dependent dust temperature is essentially important for the vertical cold finger effect.
%As most dust grains settle near the midplane in our model, far ultraviolet radiation (FUV) can %penetrate relatively deep into the disk, $z/R \sim 0.1$.
Stellar far ultraviolet (FUV) radiation starts to be attenuated at $z/R \sim 0.2$, above which the gas temperature is higher than the dust temperatures. 
FUV radiation can penetrate relatively deep into the disk;
$\chi > 1$ even at $z/R \sim 0.1$,
where $\chi$ is the wavelength-integrated FUV flux normalized by the Draine field \citep{draine78}.

\begin{figure*}[ht]
%    \centering
    \plotone{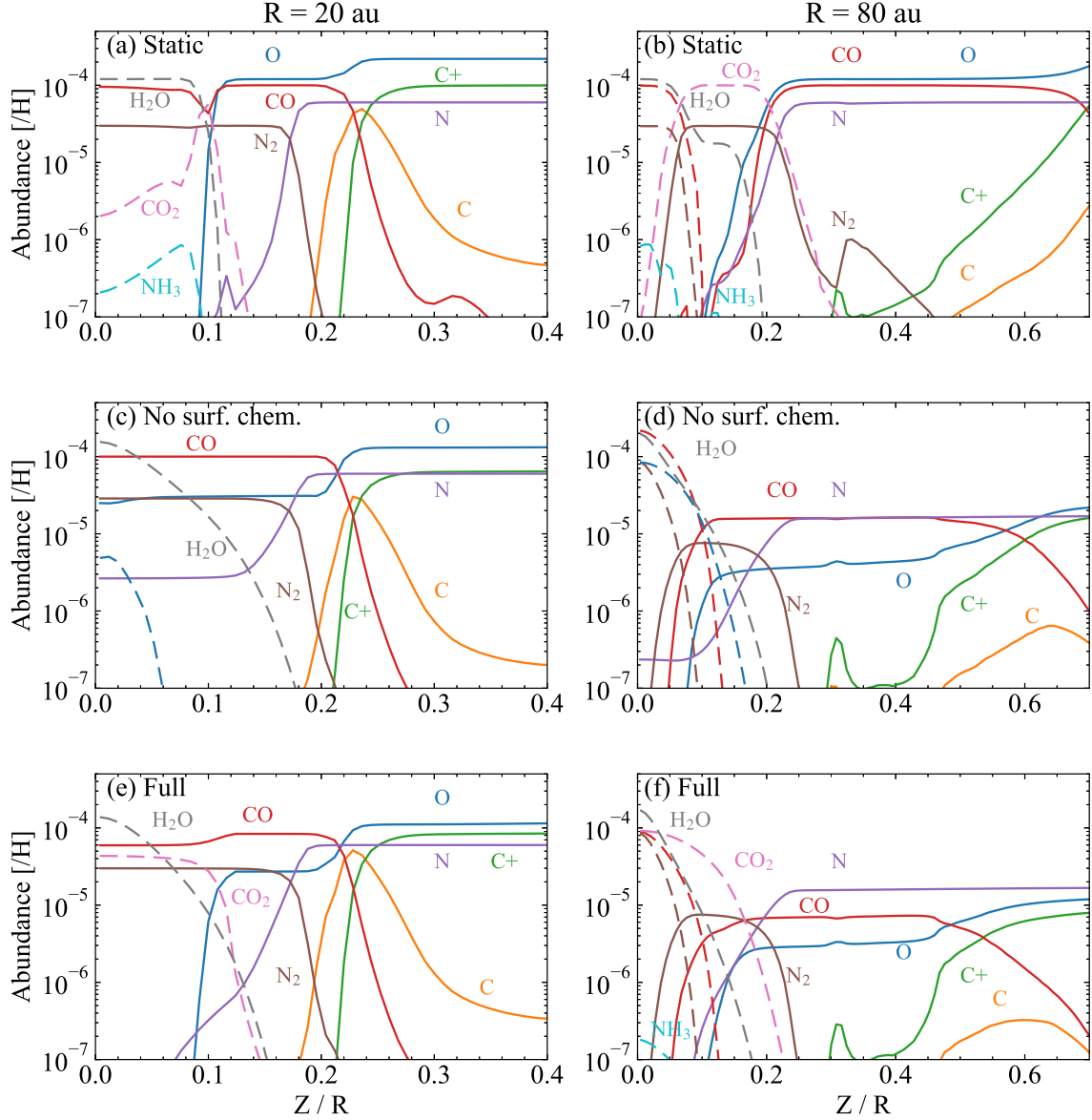}
    \caption{Vertical distributions of the abundances of selected species with respect to hydrogen nuclei at $R=20$ au (left panels) and 80 au (right panels) at 1 Myr. Top, middle, and bottom panels show the static model (i.e., only chemistry is considered), the model without grain surface chemistry, and our full model, respectively. Gas phase species are shown by solid lines, while icy species are shown by dashed lines. For icy species, the total abundances (i.e., the sum of the ice abundances on 8 dust populations) are shown.}
    \label{fig:ab}
\end{figure*}

The top panels in Figure \ref{fig:ab} show the abundances of selected species as functions of $z/R$ at $R = 20$ au and $R = 80$ au at 1 Myr in the model without any transport processes (i.e., only chemical reactions are considered; hereafter called static model).
At $R = 20$ au, gas-phase CO is abundant with the abundance of $\sim$10$^{-4}$ except at $z/R \gtrsim 0.25$, where CO is efficiently dissociated by stellar UV photons.
At $z/R \sim 0.1$, where stellar UV photons are partly attenuated ($\chi \sim 1$), \ce{CO2} ice is formed by the surface two-body reaction between CO ice and OH ice, the latter of which is formed by photodissociation of \ce{H2O} ice \citep[e.g.,][]{watanabe02,ioppolo09,oba10,noble11,eistrup16,bosman18,ruaud19,ruaud22}.
Above $z/R \sim 0.1$, the stellar UV radiation is too strong, and \ce{CO2} and \ce{H2O} ices are destroyed and not present abundantly.
The \ce{CO2}-rich regions also appear at $R = 80$ au between the CO-ice rich and CO-gas rich regions ($z/R \sim 0.1$--0.2).
The conversion of CO into \ce{CO2} is driven by stellar UV photons, and the cosmic-ray/X-ray induced UV photons are less relevant, because the flux of the latter is lower than the former by several orders of magnitude at $z/R \sim 0.1$.

The middle panels in Figure \ref{fig:ab} show the molecular abundances at 1 Myr in the model without surface chemistry but with the transport processes.
In this model, all two-body reactions on grain surfaces and photodissociation of icy species are turned off, but adsorption of gas-phase species and thermal/non-thermal desorption of icy species are considered. 
At $R = 80$ au, where the dust temperatures near the mideplane are low enough for the freeze-out of CO and \ce{N2}, the gas-phase abundances of CO and \ce{N2} become lower than their canonical values ($1\times10^{-4}$ and $3\times10^{-5}$, respectively) by  factors of $\sim$5 and $\sim$3, respectively, due to the vertical cold finger effect.
The impact of the vertical cold finger effect is more significant for CO than \ce{N2}, because of the higher binding energy of CO (1150 K) than that of \ce{N2} (1000 K).
As a result, the elemental abundances of nitrogen and carbon become similar at $R=80$ au.
%Atomic oxygen whose binding energy is set to be 1300 K is frozen-out on dust grains as it is.
At $R = 20$ au, the temperatures of all dust populations are higher than the freeze-out temperature of CO and \ce{N2} even at the midplane.
Then the cold finger effect does not work, and the gas-phase abundances of CO and \ce{N2} in the warm molecular layers are their canonical values.

The bottom panels in Figure \ref{fig:ab} show the molecular abundances at 1 Myr in our full model, where both transport processes and gas-ice chemistry are considered.
At $R = 80$ au, \ce{CO2} ice is the major reservoir of elemental carbon and oxygen 
at $z/R \lesssim 0.1$, as well as  CO ice and \ce{H2O} ice.
\ce{CO2} ice is formed by the surface reaction CO + OH $\rightarrow$ \ce{CO2} + H at $z/R \sim 0.1$--0.2, where stellar UV radiation is not significantly attenuated, and transported into the midplane, where UV radiation is completely shielded, by dust settling and turbulent mixing.
The source of oxygen for the \ce{CO2} formation is \ce{H2O} ice and atomic oxygen, the latter of which is transported from the disk upper layers. 
\ce{N2} does not easily react with radical species on the grain surface in contrast to CO.
Owing to the much larger binding energy of \ce{CO2} (and \ce{H2O}) compared to that of \ce{N2},
elemental carbon and oxygen are easily locked up as ices near the midplane than nitrogen.
Consequently, nitrogen becomes the third most abundant element in the gas phase after hydrogen and helium at $R=80$ au.

At $R = 20$ au, inside the CO snowline, \ce{CO2} ice is formed at $z/R \sim 0.1$ by the surface reaction CO + OH $\rightarrow$ \ce{CO2} + H, and transported into the midplane by dust settling and turbulent mixing like at $R = 80$ au.
Because of the \ce{CO2} formation, the gas-phase CO abundance becomes lower than the canonical value by a factor of $\sim$2 at 1 Myr.
The abundance of \ce{CO2} ice remains lower than those of CO gas and \ce{H2O} ice at $R = 20$ au.
The main limiting factor of the \ce{CO2} formation is the amount of available oxygen for the chemistry.
At 80 au (outside the CO snowline), atomic O and \ce{H2O} ice at $z/R \gtrsim 0.1$ are abundant enough to convert the significant fraction of CO gas into \ce{CO2} ice, 
because CO near the midplane is frozen out onto dust grains, and thus abundant only in the gas phase at $z/R \gtrsim 0.1$.
\ce{H2O} ice below $z/R \lesssim 0.1$ does not contribute to the \ce{CO2} formation, 
because \ce{H2O} ice is mostly on large ($\gtrsim$100 $\mu$m) dust grains, which are dynamically decoupled from the gas, 
and because the stellar UV radiation to dissociate \ce{H2O} ice is significantly attenuated.
At 20 au (inside the CO snowline), however, atomic O and \ce{H2O} ice at $z/R \gtrsim 0.1$ are not abundant enough, 
because CO gas is present throughout (both in the surface layer and at the midplane).
As a result, the conversion of CO gas into \ce{CO2} ice is not very efficient inside the CO snowline compared to outside the CO snowline.

Our models neglect radial drift of dust grains.
As discussed in Appendix \ref{append:ice}, the size of dust grains which contributes to the grain surface chemistry efficiently is $\lesssim$100 $\mu$m, 
and larger dust grains do not play a significant role on the chemistry.
The dimensionless stopping time of 100 $\mu$m-sized dust grains is $\sim$10$^{-3}$ and $\sim$5$\times10^{-3}$ at the midplane of $R=20$ au and 80 au, respectively.
For a 100 $\mu$m-sized dust grain, the timescale of drifting from 80 au to 30 au (i.e., around the CO snowline) is evaluated to be $\gtrsim$1 Myr using Eqs. 17 and 18 in \citet{brauer08}.
Then the impact of radial drift on our results on the carbon and nitrogen depletion outside the CO snowline would be not significant.
Inside the CO snowline, radial drift of CO-ice mantled dust grains from the outer regions, followed by the sublimation of CO ice can enhance the CO gas abundance \citep[e.g.,][]{krijt20}.
This effect would not be negligible, as not small amount of CO is present on large dust grains ($\gtrsim$100 $\mu$m).

\subsection{Depletion factors of carbon and nitrogen}

\begin{figure*}[th]
%    \centering
    \plotone{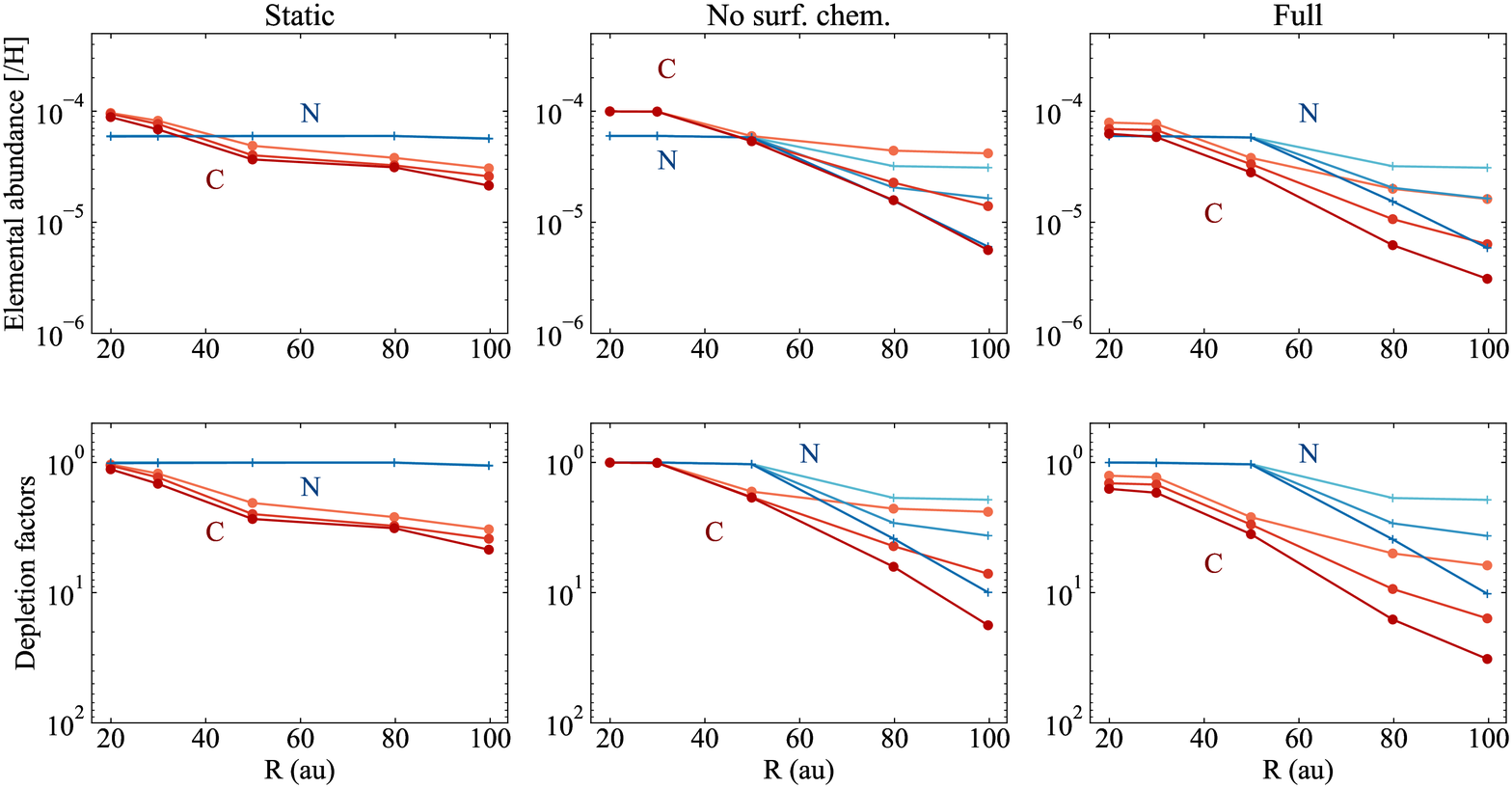}
    \caption{Abundances (top) and depletion factors (bottom) of elemental carbon (red lines) and nitrogen (blue lines) above the snow surfaces of CO and \ce{N2}, respectively, in the static  model (left), the model without grain surface chemistry (middle), and in the full model (right) at 0.1 Myr, 0.3 Myr, 
    and 1 Myr. The depletion factors monotonically increase with time.}
    \label{fig:dep_fac}
\end{figure*}

We define depletion factors of element X as 
%\begin{linenomath*}
\begin{align}
\fdep{X}= [{\rm X/H}]_0/[{\rm X/H}]_{ss},
\end{align}
%\end{linenomath*}
where $[{\rm X/H}]_0$ is the initial elemental abundance of X with respect to H, and $[{\rm X/H}]_{ss}$ is the elemental abundance of X with respect to H in the gas above the snow surfaces of CO for carbon and \ce{N2} for nitrogen \citep[cf.][]{zhang19,krijt20}.
Larger value of $\fdep{X}$ means that the depletion of element X in regions above the snow surface is more significant.
We define the position of the CO snow surface as the height below which multilayered CO ice can be present on at least one of the 8 dust populations.
We compare the total desorption rate of CO by thermal desorption and photodesorption with its adsorption rate for each dust population, 
assuming the gas-phase CO abundance is 10$^{-4}$ and grain surfaces 
are covered by one-monolayer of CO ice.
The snow surface of CO is defined as the height above which the desorption rate of CO is greater than the adsorption rate for all dust populations.
Similar definition is applied to the snow surface of \ce{N2}.
The positions of the snow surfaces are the same among different models as long as we adopt the same disk physical model.

Figure \ref{fig:dep_fac} shows $\fdep{C}$ and $\fdep{N}$ as functions of $R$ in the static model (left panel), the model without surface chemistry (middle panel), and our full model (right model) at $t = 0.1$ Myr, 0.3 Myr, and 1 Myr.
As we ran our models only at $R$ = 20, 30, 50, 80, and 100 au, we cannot infer the exact radial positions of the CO and \ce{N2} snowlines; the CO snowline is located at around 30 au, 
while the \ce{N2} snowline is located at around 50 au.
In the static model, $\fdep{N}$ is almost unity (i.e., no significant depletion) regardless of $R$, while $\fdep{C}$ is $\sim$3--5 well outside the CO snowline.
%As a result, $\fdep{C}/\fdep{N}$ is around 8 well outside the CO snowline.
In the model without surface chemistry, $\fdep{C}$ is in the range of $\sim$2--20 outside the CO snowline, while it is close to unity inside the CO snowline. 
At $R = 50$ au, the CO snow surface is close to the midplane compared to that at $R \ge 80$ au, leading to the lower efficiency of the vertical cold finger effect.
The radial profile of $\fdep{N}$ basically follows that of $\fdep{C}$, and $\fdep{C}/\fdep{N}$ is $\sim$1--2.
In the full model, $\fdep{N}$ is similar to that in the model without surface chemistry, because nitrogen in the warm gas is depleted only by the cold finger effect of \ce{N2}.
%, and because the surface chemistry does not play any significant roles.
$\fdep{C}$ in the full model is the largest ($\sim$3--30) among the three models, because of the combination of the vertical cold finger effect and the surface chemistry which converts CO into \ce{CO2}.
Consequently, $\fdep{C}/\fdep{N}$ can be as high as $\sim$3--5 in the full model, i.e., carbon is more depleted than nitrogen by a factor of $\gtrsim$3 outside the CO snowline, leading to higher abundance of gas-phase nitrogen than that of gas-phase carbon.
We also run some additional models to explore the parameter dependence of $\fdep{C}$ and $\fdep{N}$, varying $\xi_{\rm CR}$, $a_{\rm max}$, and $a_{\rm min}$ (see the Appendix \ref{append:param} in details).
We confirmed that these parameters did not affect our qualitative chemical results.
%We find that $\fdep{C}$ is larger than $\fdep{N}$, regardless of the values of those parameters.
Then, the combination of the cold finger effect and the grain surface chemistry results in larger depletion of carbon than nitrogen outside the CO snowline within 1 Myr once the dust grains have settled towards the midplane, making nitrogen the third most abundant element in the gas phase after hydrogen and helium.
Our full model is consistent with the observations toward the IM Lup disk with the age of 0.5--1 Myr,
where carbon and oxygen are depleted compared to the ISM abundance by a factor of $\sim$20 or even larger, whereas nitrogen depletion is less significant \citep{cleeves18}.

%\begin{figure*}[t]
%    \centering
%    \plotone{100au_amax_elem.eps}
%    \caption{Spatial distributions of the C/H ratio (a), O/H ratio (b), the N/H ratio (c), the C/O ratio (d), and the C/N ratio (e). The total (gas+ice) elemental abundance ratios are shown by solid lines, while the gas-phase elemental abundance ratios are shown by dashed lines. Different colors in each panel represents different time from 0.1 Myr to 3 Myr.}
%    \label{fig:diff_dist}
%\end{figure*}

\section{Discussion} \label{sec:discuss}
\subsection{Timescale of elemental depletion: UV driven chemistry versus CR driven chemistry} \label{sec:timescale}
In our full model, outside the midplane CO snowline, the volatile carbon abundance in the warm molecular layer can be reduced by a factor of more than 10 within 1 Myr due to the combination of the vertical cold finger effect and the chemical conversion of CO into \ce{CO2} ice driven by stellar UV photons.
On the other hand, the nitrogen abundance in the warm molecular layer is reduced solely due to the vertical cold finger effect of \ce{N2}.
%Then, nitrogen can become more abundant than carbon in the warm molecular layer within 1 Myr.
Here we discuss which processes determine the evolutionary timescale of the elemental abundances in the warm molecular layer.

When the turbulent mixing is switched off (the static model), the chemical conversion of CO into \ce{CO2} ice occurs only in the limited regions ($z/R \sim 0.1$--0.2).
In the full model, however, \ce{CO2} ice formed at $z/R \sim 0.1$--0.2 can be transported to midplane via the turbulent mixing and dust settling, 
while CO near the midplane and in the disk upper layers can be transported to the regions at $z/R \sim 0.1$--0.2 via the turbulent mixing.
Then, the \ce{CO2}-rich regions are expanded with time.
The timescale of this process is limited by mixing, because the timescale of the freeze-out of CO and atomic O onto dust grains ($\tau_{\rm fr}$) and that of the chemistry driven by stellar UV photons ($\tau_{\rm phd}$) are shorter than the mixing timescale.
The turbulent mixing timescale is given by $\tau_{\rm mix} \sim (0.1 r)^2/D_z$.
For example, $\tau_{\rm mix}$ is $\sim$3$\times10^4$ yr at $R = 20$ au, $\sim$8$\times10^4$ yr at 50 au, and $\sim$2$\times10^5$ yr at 100 au in our models, where $\alpha$ is assumed to be 10$^{-3}$.
The freeze-out timescale $\tau_{\rm fr}$ is given by
%\begin{linenomath*}
\begin{equation}
\tau_{\rm fr} \approx 3000 \left(\frac{10^{-24} \, {\rm cm^2/H}}{\sigma_d} \right) \left(\frac{10^9\, {\rm cm^{-3}}}{n_{\rm H}} \right)  \,\,{\rm yr},
\end{equation}
%\end{linenomath*}
where $\sigma_d$ is the total dust cross section per hydrogen nuclei, and $n_{\rm H}$ is the number density of hydrogen nuclei.
As OH ice is produced by \ce{H2O} ice photodissociation, $\tau_{\rm phd}$ is given by
%\begin{linenomath*}
\begin{equation}
\tau_{\rm phd} \approx 100 (\chi/1)^{-1} \,\,{\rm yr}.
\end{equation}
%\end{linenomath*}
Cosmic-ray/X-ray induced UV photons are less important than the stellar UV;
the FUV flux induced by cosmic rays in the dense molecular cloud core environment, where $\xi$ is $\sim$10$^{-17}$ s$^{-1}$, is $\sim$10$^4$ photons cm$^{-2}$ s$^{-1}$.
As $\chi = 1$ corresponds to the FUV flux of $\sim$10$^8$ photons cm$^{-2}$ s$^{-1}$,
the cosmic-ray/X-ray induced UV photons are negligible unless $\xi \gtrsim 10^{-14}$ s$^{-1}$ (see Figure \ref{fig:phys}).
Then $\tau_{\rm mix}$ is much shorter than $\tau_{\rm fr}$ and $\tau_{\rm phd}$.

On the other hand, there are many previous modeling studies which show that CO is converted into less volatile species via the chemistry driven by cosmic-rays and X-rays \citep[e.g.,][]{bergin14,furuya14}.
In brief, CO is destroyed by \ce{He+} to produce \ce{C+} and atomic O, which are eventually locked up in icy \ce{CO2}, icy \ce{CH4}, and icy carbon-chain molecules.
As these molecules are much less volatile than CO, they are present in ice mantles even inside the CO snowline.
As the rate-limiting step of the CO gas reprocessing chemistry is the ionization of \ce{He} by cosmic-rays/X-rays to produce \ce{He+}, the timescale is given by 
%\begin{linenomath*}
\begin{align}
\tau_{\rm CRchem} &\approx \frac{x_{\rm CO}}{\xi_{\rm He} x_{\rm He}} \nonumber \\
                  &\approx 2 \times 10^{7}\left(\frac{x_{\rm CO}}{1\times 10^{-4}}\right)
\left(\frac{\xi_{\rm He}}{5 \times 10^{-19}\,{\rm s^{-1}}}\right)^{-1} \,\,{\rm yr}, \label{coms:eq:t_co}
\end{align}
%\end{linenomath*}
where $x_i$ is the abundance of species $i$, and ${\xi_{\rm He}}$ is the ionization rate of He atom \citep{furuya14}, which is a half of the ionization rate of \ce{H2}.
Then $\tau_{\rm CRchem}$ is much larger than $\tau_{\rm phd}$ and $\tau_{\rm mix}$.
\ce{N2} can also be converted into less volatile species (\ce{NH3} ice) by the following pathway \citep{willacy07,furuya14}: 
\ce{N2} $\xrightarrow{\ce{H3+}}$ \ce{N2H+} $\xrightarrow{{\rm e^-}}$ NH. 
Once NH is formed, it is adsorbed onto dust grains, followed by subsequent hydrogenation to form \ce{NH3} ice.
The timescale of the \ce{N2} conversion into \ce{NH3} ice is slightly longer than $\tau_{\rm CRchem}$, as it occurs only after the significant depletion of CO gas (when the CO abundance is lower than $\sim$10$^{-5}$); otherwise \ce{N2H+} is destroyed by CO and \ce{N2} is reformed \citep{furuya14}.
$\tau_{\rm CRchem}$ is much longer than $\tau_{\rm phd}$ and $\tau_{\rm mix}$, and thus the chemistry driven by cosmic-rays (and X-rays) are less important than that driven by stellar UV photons.
Even if the higher value for $\xi_{\rm CR}$ of 10$^{-16}$ s$^{-1}$ is assumed, the above conclusion holds, 
because $\tau_{\rm CRchem}$ is still larger than  $\tau_{\rm fr}$, $\tau_{\rm phd}$, and $\tau_{\rm mix}$.
Indeed, the depletion factors of carbon and nitrogen do not change significantly between the model with $\xi_{\rm CR} = 10^{-18}$ s$^{-1}$ and that with $\xi_{\rm CR} = 10^{-16}$ s$^{-1}$ (see Appendix \ref{append:param}).

%Therefore, once the dust grains have grown and settled toward the midplane, triggering the vertical cold finger effect and deeper penetration of stellar UV photons into disks, orders of magnitude depletion of elemental carbon can occur within 1 Myr when $\alpha = 10^{-3}$.
%The depletion of nitrogen can occur on a similar timescale, but the degree of nitrogen depletion is much smaller than that of carbon and oxygen.

\subsection{Dependence on the strength of turbulence}
\begin{figure*}[ht]
%    \centering
    \plotone{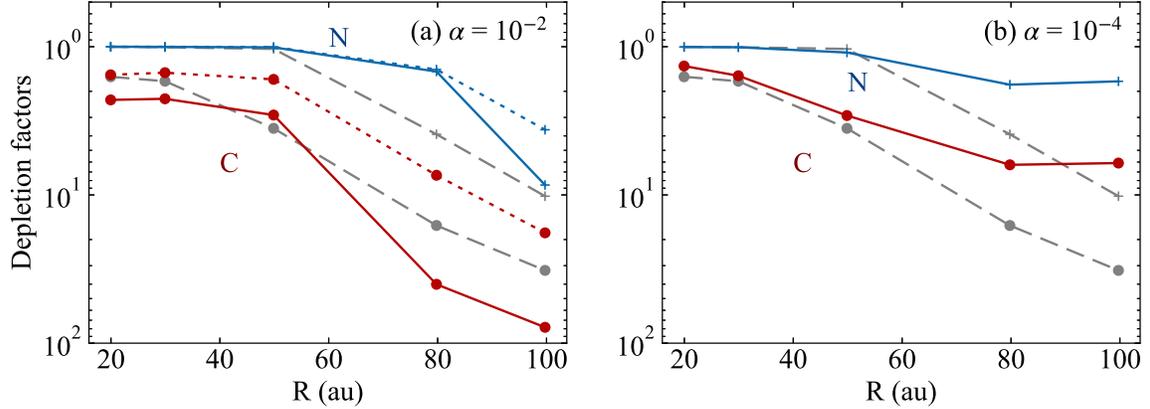}
    \caption{Depletion factors of elemental carbon (red solid lines) and nitrogen (blue solid lines) in the full model with $\alpha = 10^{-2}$ (left) and with $\alpha = 10^{-4}$ (right) at 1 Myr. Dashed gray lines represent the depletion factors in the model with $\alpha = 10^{-3}$ at 1 Myr for comparisons.
    The dotted lines in the left panel shows the depletion factors in the model with $\alpha = 10^{-2}$ at 0.1 Myr.
    }
    \label{fig:dep_fac_mix}
\end{figure*}

In our models presented so far, the $\alpha$ parameter for the vertical diffusion coefficient is assumed to be 10$^{-3}$.
As discussed in Section \ref{sec:timescale}, the timescale of carbon and nitrogen depletion is mostly determined by the turbulent mixing timescale, and thus the $\alpha$ parameter should affect the timescale and the degrees of the carbon and nitrogen depletion at given time.
Recent observational studies have put constraints on the strength of the turbulence in the warm molecular layers for some disks on the basis of the molecular line observations.
\citet{flaherty20} constrained the non-thermal gas motion ($\delta v$) in disks around DM Tau, MWC 480, and V4046 Sgr, based on the \ce{^12CO} 2--1 observations and parametric disk models. 
Converting $\delta v$ to $\alpha$ by $\alpha = (\delta v/c_s)^2$,
they found that $\alpha \sim 8 \times 10^{-2}$ for the DM Tau disk, 
$\alpha < 3 \times 10^{-3}$ for the MWC 480 disk, 
and $\alpha < 1 \times 10^{-2}$ for the V4046 Sgr disk in the gas traced by \ce{^12CO} 2--1, i.e., not the disk midplane.
Similar observational studies toward the disks around TW Hya and HD 162936 indicate $\alpha < 7 \times 10^{-3}$ and $\alpha < 3 \times 10^{-3}$, respectively \citep{teague16,flaherty17,flaherty18}.
%The value of $\alpha$, in particular, close to the midplane can be constrained from dust continuum observations, because the degree of dust settling depends on the strength of the turbulence.
%\citet{pinte16} concluded that the HL Tau disk shows the substantial dust settling and $\alpha$ is $\sim$10$^{-4}$ from the morphology of dust gaps.
%\citet{doi21} constrained the dust scale height in dust rings of the HD 163296 disk from the azimuthal intensity variation of the rings.
%If the dust size of 1 mm is assumed, $\alpha$ is constrained to be $>$2 $\times$ 10$^{-2}$ at the inner ring (68 au), 
%and $<$2 $\times$ 10$^{-4}$ at the outer ring (100 au).

These observational studies seem to indicate that the strength of the turbulence is not strong ($\alpha \lesssim 10^{-3}$) in disks in general, but the turbulence is strong ($\alpha \gtrsim 10^{-2}$) in some disks.
To explore the impact of the $\alpha$ value on the model results, we recalculated the entire modeling processes (Figure \ref{fig:flowchart}) with $\alpha = 10^{-2}$ and $\alpha = 10^{-4}$ as the value of $\alpha$ parameter also affects the disk physical conditions.
Figure \ref{fig:dep_fac_mix} shows $\fdep{C}$ and $\fdep{N}$ in full models with $\alpha = 10^{-2}$ (left panel) and $\alpha = 10^{-4}$ (right panel).
As the mixing timescale is inversely proportional to $\alpha$, $\fdep{C}$ already reaches $\sim$10 at 0.1 Myr at $R \ge$ 80 au in the model with $\alpha = 10^{-2}$.
Both $\fdep{C}$ and $\fdep{N}$ reach the steady state values within 1 Myr in the model with $\alpha = 10^{-2}$.
In the model with $\alpha = 10^{-3}$, $\fdep{N}$ almost reaches the steady state value within 1 Myr, 
while $\fdep{C}$ does not.
We reran the model with $\alpha = 10^{-3}$ for 3 Myr, 
and confirmed that $\fdep{C}$ reaches $\sim$50 and $\sim$60 at $R=80$ au and $R=100$ au, respectively, at 3 Myr,
while $\fdep{C}$ at $R \le 50$ au almost reaches the steady state value within 1 Myr.
%The depletion factors in the models with $\alpha = 10^{-2}$ and $\alpha = 10^{-3}$ at 1 Myr are also similar, indicating that the steady state values of the depletion factors are similar in the two models.
In the model with $\alpha = 10^{-4}$, the depletion factors are less than 10 even at 1 Myr, because of the long mixing timescale (e.g., longer than 1 Myr at $R \gtrsim 50$ au).
Therefore, relatively large value of $\alpha$ ($\gtrsim$10$^{-3}$) is favorable for the significant depletion of carbon and making the warm molecular layer enriched in nitrogen relative to carbon within 1 Myr.

On the other hand, \citet{anderson19} found that \ce{N2H+}/\ce{CO} flux ratio is higher in 5-10 Myr disks than younger disks with the age of $<$2 Myr.
If this trend primary traces the evolution of volatile elemental abundances,
the depletion of carbon should continue in 5-10 Myr.
Because time it takes for the depletion factors reach steady state is shorter for higher $\alpha$ value, our model with $\alpha$ = 10$^{-4}$ or 10$^{-3}$ 
may be more consistent with the finding by \citet{anderson19} rather than the model with $\alpha = 10^{-2}$.
Taken together, to explain the observed CO (or carbon) depletion timescale by our models, the reasonable value of $\alpha$ is $\sim$10$^{-3}$. 

%$\alpha$ can depend on the disk position.

%In our models, the value of $alpha$ is fixed with time.
%It should be noted, however, that if $\alpha$ was large in the past, carbon depletion occurs quickly, 
%after depletion occurs, the depletion would remain even if $\alpha$ becomes lower.
%Then,  evolution is important.

\subsection{Column densities of gas-phase molecules} \label{sec:column}
\begin{figure*}[ht]
%    \centering
    \plotone{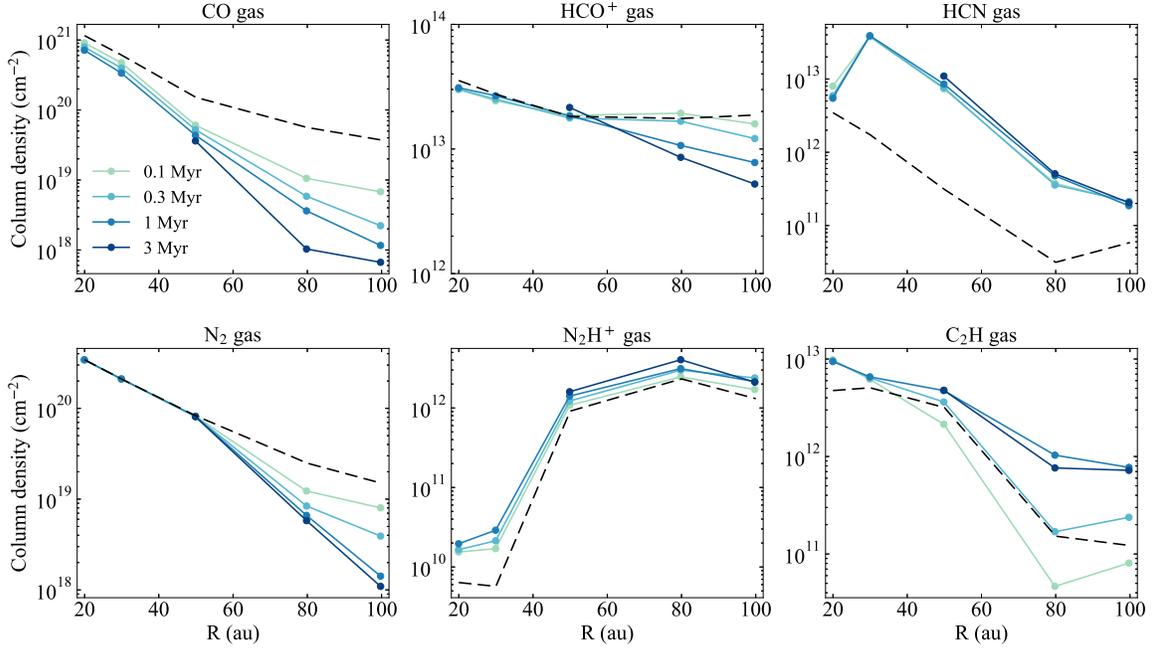}
    \caption{
    Radial profiles of the column densities of selected gas-phase molecules in the full model at 0.1 Myr, 0.3 Myr, 1 Myr, and 3 Myr. The column densities at 3 Myr are shown only 
    at $R$ = 50, 80, 100 au. At the inner radii, the column density almost reaches the steady state values within 1 Myr.
    For comparisons, black dashed lines show the column densities predicted by the static model without grain surface chemistry (i.e., no depletion of carbon and nitrogen in the warm molecular layers) at 1 Myr.}
    \label{fig:mol_column}
\end{figure*}

Here we discuss how the column densities of gas-phase species evolve with time in our full model with $\alpha = 10^{-3}$.
Figure \ref{fig:mol_column} shows the column densities of CO and \ce{N2}, their protonated species (\ce{HCO+} and \ce{N2H+}), 
and the prove of the elemental carbon-to-oxygen ratio in the gas-phase (hereafter $\cogas$; \ce{C2H} and HCN) at $t = 0.1$, 0.3, 1 Myr, and 3 Myr.
For comparisons, we ran an additional static model, where grain surface chemistry, except for desorption and adsorption processes, was turned off (i.e., there is no depletion of carbon, oxygen, and nitrogen in the warm molecular layers).
The molecular column densities predicted by this additional model at 1 Myr is shown by black dashed lines in the figure.

The column densities of CO and \ce{HCO+} decrease with time, reflecting the gradual loss of CO from the warm molecular layer via the combination of the chemical conversion of CO and the vertical cold finger effect.
The \ce{N2} column density decreases with time at $>$50 au, due to the vertical cold finger effect only.
The column density of \ce{N2} is larger than that of CO, indicating that \ce{N2} is the second most abundant molecule in the gas-phase.
The column densities of \ce{N2} decrease with time, 
whereas the \ce{N2H+} column density increases slightly with time rather than decreased.
As \ce{N2H+} is destroyed by CO, the decrease of the gas-phase CO abundance leads to the enhanced \ce{N2H+} abundance \citep[e.g.,][]{aikawa15}, canceling the decrease of the gas-phase \ce{N2} abundance.
As a result, the column density ratio of CO to \ce{N2H+} increases with time at least until 3 Myr.
This result may be consistent with the observations by \citet{anderson19}, 
who found that the \ce{N2H+}/CO flux ratio in 5–11 Myr old disks in Upper Sco is higher than that in younger disks with the age of $<$2 Myr.
\citet{trapman22} showed that the combination of \ce{C^18O} and \ce{N2H+} lines is useful to constrain disk gas mass, adopting a static thermal-chemical disk model, assuming that nitrogen is not depleted.
Our results do not fully support their assumption on the non-depletion of nitrogen.
However, combination of \ce{C^18O} and \ce{N2H+} lines would be useful to constrain the disk gas mass, because the carbon depletion is more significant than the nitrogen depletion, 
and thus the \ce{N2H+}/CO ratio can be used as a probe of the carbon (CO) depletion.

\ce{C2H} is considered as a prove of $\cogas$, because its abundance depends on whether $\cogas$ is higher than unity or not \citep[e.g.,][]{bergin16,miotello17,bosman21}.
In our full model, the \ce{C2H} column density increases with time by a factor of $\sim$10 at most,
because $\cogas$ becomes close to, but still lower than unity both inside and outside of the CO snowline (see panels e and f of Figure \ref{fig:ab}).
The HCN column density remains almost unchanged with time in our full model.
%In our models, as \ce{H2O} ice is locked up in large dust grains, [C/O] of materials available for the gas-ice chemistry is $\sim$1 in our models.
%The formation of \ce{CO2} ice does not lead to higher [C/O] than unity, because stellar UV dissociating CO is attenuated in the \ce{CO2} forming regions. 
The \ce{C2H} and HCN column densities in our model ($\sim$10$^{12}$--10$^{13}$ cm$^{-2}$ and $\sim$10$^{11}$--10$^{13}$ cm$^{-2}$, respectively) at 1 Myr are still one or two orders of magnitude lower than that observed in the outer regions (outside of the CO snowline) of Class II disks \citep{bergin16,bergner20,guzman21}.
The underestimation of the \ce{C2H} and HCN column densities in our model indicates a need for additional process to make $\cogas$ higher than unity.

The rationale for $0.5 \lesssim \cogas \lesssim 1$ both inside and outside of the CO snow line in our model is as follows.
Inside the CO snowline, only the oxygen-rich species, \ce{H2O} and \ce{CO2}, can freeze out near the midplane.
Then, the gas becomes dominated by CO over time, leading to $\cogas \sim 1$.
Outside the CO snowline, there are two CO reservoirs at $t = 0$ yr; one is the gas-phase CO in the warm molecular layer and another is the CO ice on small dust grains near the midplane. 
As CO is assumed to be present in the gas phase at $t = 0$ yr, it is preferentially frozen-out onto smaller grains;
smaller grains dominate dust cross section, and dust temperature does not depend on the size near the midplane.
Some CO ice near the midplane is transported upward, enriching the warm molecular layers in CO.
Atomic O and CO in the surface layer are consumed by the \ce{CO2} formation at a similar rate, 
but this enrichment process partially cancels out the CO depletion in the surface layers.
As a result, $\cogas$ becomes $>$0.5, but does not exceed unity.

A promising process for enhancing $\cogas$ is the destruction of carbon grains by photoablation as proposed by \citet{bosman21b}.
The carbon abundance locked up in refractory carbon dust is $10^{-4} \times 100 \Delta_{\rm d/g}$, where $\Delta_{\rm d/g}$ is the dust-to-gas mass ratio. 
To enhance $\cogas$ considerably ($\sim$2), the carbon abundance released in the gas phase should be comparable to or larger than the gas-phase carbon abundance.
Then the following relation should hold, $\fdep{C}\Delta_{\rm d/g} \gtrsim 10^{-2}$ \citep{bosman21b}, assuming complete photoablation of refractory carbon dust.
In our model with $\alpha = 10^{-3}$ at 1 Myr, the requirement is satisfied at $z/R \lesssim 0.2$, even if complete photoablation of carbon grains is assumed.
Below $z/R \lesssim 0.1$, however, stellar UV is significantly attenuated ($\chi \ll 1$) and/or the dust temperature is too low leading to the freeze-out of gas-phase carbon.
Taken together, only very limited regions ($z/R \sim 0.1$--0.2) might be able to become $\cogas > 1$ by the photoablation.
Even if so, locally high $\cogas$ would be smeared out by turbulent mixing in relatively short timescale (e.g., $\tau_{\rm mix} \sim 2\times 10^5$ yr at $R = 100$ au).
Therefore, as noted in \citet{vanclepper22}, carbon grain destruction may be ongoing over the course of grain growth, rather than solely after the grain growth. 

\subsection{Comparisons with previous studies} \label{sec:compare}
\citet{krijt20} studied the depletion of carbon and oxygen combining the dust evolution and grain surface chemistry of carbon and oxygen driven by cosmic rays, but ignored the chemistry driven by stellar UV photons.
%\citet{vanclepper22} combined a toy model of the pebble formation and a gas-ice astrochemical model (but surface chemistry was ignored except for hydrogenation, i.e., the \ce{CO2} formation does not occur in their model).
%The preferential loss of \ce{H2O} from the disk upper layers together with small dust grains also occurs in the model by \citet{vanclepper22}, although the vertical settling of pebbles are not explicitly included in their model.
One of the most significant differences between their study and ours is that they considered the formation of pebbles from small grains in disks, 
while we assumed that pebbles already exist at the beginning of our simulations.
In \citet{krijt20}, the formation and the vertical settling of pebbles transport \ce{H2O} ice to the midplane.
This process leads to $\cogas \sim 1$, because CO is much more volatile than \ce{H2O}.
On the other hand, in our models, we assumed that pebbles already exist at $t = 0$ yr and the ISM-like elemental abundance throughout the disk 
(i.e., $\cogas \sim 0.5$ in the disk upper layers, where \ce{H2O} is efficiently dissociated by stellar UV photons).
As noted in Section \ref{sec:init}, what happens to water in young disks, where the growth of small dust grains have already started, remains unclear.
The model by \citet{krijt20} would correspond to the scenario 
where the sublimation/photodesorption of water ice is negligible, and coagulation and settling of dust covered by water ice cause the depletion of volatile oxygen in the disk upper layers,
because their model did not consider the chemistry induced by stellar UV and the disk is cold.
Our model corresponds to another extreme scenario where water ice on dust grains are uncovered due to thermal desorption and/or photodesorption of water ice in the disk upper layers, 
and dust coagulation and settling do not cause the depletion of volatile oxygen in the disk upper layers.
Reality would be somewhere between the two extremes, depending on stellar properties and disk physical structures.
Note that the oxygen abundance in the disk upper layers would also be related to the photoablation of carbon grain; 
if the photoablation occurs efficiently during the formation and settling of pebbles, 
\ce{H2O} ice, which covers dust grains, would be photodesorbed at the same time.

\begin{figure}[ht]
%    \centering
    \plotone{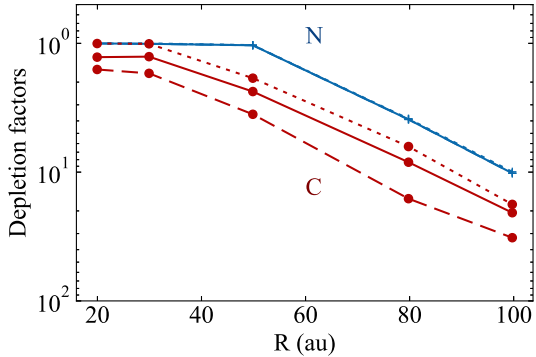}
    \caption{
    The depletion factors of elemental carbon (red solid line) and nitrogen (blue solid line) in the model where the initial \ce{H2O}-to-dust mass ratio is uniform inside the disk at 1 Myr. For comparisons, dashed lines and dotted lines represent the depletion factors in the fiducial full model and the model without surface chemistry. In these two models, the initial \ce{H2O} abundance is uniformly 10$^{-4}$.
    }
    \label{fig:dep_fac_odep}
\end{figure}

In order to check the impact of elemental oxygen abundance available for the \ce{CO2} formation on our results, 
we reran our full model assuming that the initial \ce{H2O} abundance scales with the local dust-to-gas mass ratio, i.e., 10$^{-4} \times (\Delta_{\rm d/g}/10^{-2})$.
In other words, the initial \ce{H2O}-to-dust mass ratio is uniform inside the disk.
As $\Delta_{\rm d/g}$ is $\sim$10$^{-3}$ or lower at $z/R \gtrsim 0.1$ (see Fig. \ref{fig:phys}), the elemental abundance of oxygen not locked up in CO is lower than that in our fiducial models by a factor of $\sim$10 or even larger.
We confirmed that $\fdep{C}$ in the model with the uniform \ce{H2O}-to-dust mass ratio is between that in the fiducial full model and that in the model without grain surface chemistry (Figure \ref{fig:dep_fac_odep}).
We also confirmed that $\cogas$ at 1 Myr is close to unity as in the fiducial full model. 
Therefore, our qualitative chemical results are not sensitive to the initial oxygen abundance.
%The results imply that the quantitative difference between $\fdep{C}$ and $\fdep{N}$ in Class II disks could be used to prove how much elemental oxygen is pulled down to the midplane when the formation and the vertical settling of pebbles occurred in Class 0/I phases.
%It should be noted, however, other parameters, such as $a_{\rm max}$, temperature structure, and photoablation of carbon dust, also affect the $\fdep{C}/\fdep{N}$.

Regarding to $\cogas$ at 1 Myr or later, our model predicted that $\cogas$ becomes larger than the initial value of $\sim$0.5, but does not exceed unity.
\citet{vanclepper22}, who combined a toy model of the pebble formation and a gas-ice astrochemical model (but surface chemistry was ignored except for hydrogenation, i.e., the \ce{CO2} formation does not occur in their model), also predicted that $\cogas$ does not exceed unity.
On the other hand, \citet{krijt20} predicted that eventually $\cogas$ in the warm molecular layers can become larger than unity.
The difference in the predicted $\cogas$ can be explained by the cosmic-ray driven chemistry versus UV driven chemistry.
The rationale why $\cogas > 1$ can be realized in \citet{krijt20} is that they assumed \ce{C+}, which is produced by the destruction of CO by \ce{He+}, is eventually locked up in \ce{CH4}, while atomic O is locked up in \ce{H2O} and/or \ce{CO2}. 
As \ce{CH4} is more volatile than \ce{H2O} and \ce{CO2}, it can be more easily go back to the gas phase and enhance $\cogas$ in the warm molecular layer.
In the model by \citet{krijt20}, the effect of stellar UV on chemistry was neglected in contrast to our models.
More comprehensive simulation with dust settling, growth, and photoablation with both CR and UV driven chemistry are required 
for better understanding of the evolution of carbon, nitrogen, and oxygen in protoplanetary disks.

\subsection{Ice compositions in the midplane}
Although the main topic of this study is the gas-phase elemental abundance in the warm molecular layers, we discuss ice compositions in the disk midplane briefly, because the chemical compositions in the warm molecular layer and those in the midplane are related to each other through the turbulent mixing.
We only focus on the most abundant species in the ISM ice and cometary ice, i.e., \ce{H2O}, CO, and \ce{CO2} for C- and O-bearing species, and \ce{N2} and \ce{NH3} for N-bearing species.

\begin{figure*}[ht]
%    \centering
    \plotone{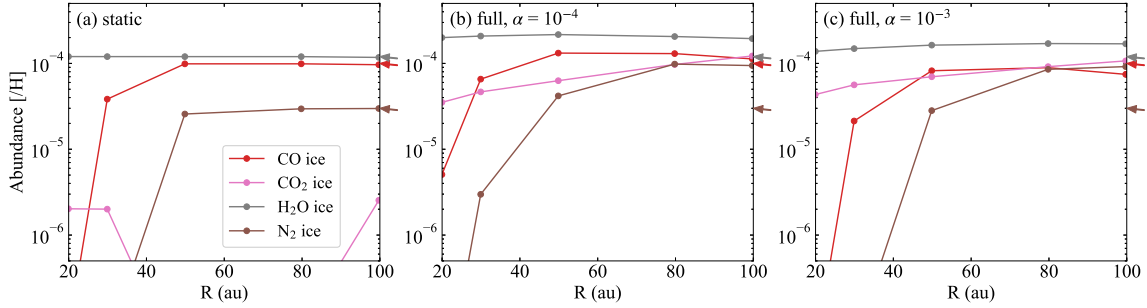}
    \caption{Radial profiles of the total abundances of selected icy molecules (i.e., the sum of the ice abundances on 8 dust populations) in the static model (left), in the full model with $\alpha = 10^{-4}$ (middle), and in the full model with $\alpha = 10^{-3}$ (right) at 1 Myr.
    Arrows on the right-hand margin indicate the initial abundances of \ce{H2O}, CO, and \ce{N2}. The abundance of \ce{NH3} is below 10$^{-6}$ and not shown in the figure.}
    \label{fig:iceab_mid}
\end{figure*}

Figure \ref{fig:iceab_mid} shows the total abundances of selected icy molecules (i.e., the sum of the ice abundances on 8 dust populations) in the disk midplane in the static model (left), in the full model with $\alpha = 10^{-4}$ (middle), and in the full model with $\alpha = 10^{-3}$ (right) at 1 Myr.
In the static model, the ice composition in the midplane is mostly determined by the assumed initial compositions and the balance between adsorption and desorption, because the timescale of the chemistry driven by cosmic-rays is longer than 1 Myr for $\xi_{\rm CR} \lesssim 10^{-17}$ s$^{-1}$ \citep[e.g.,][]{furuya14}.
When the turbulent mixing is considered, we see the increases of the \ce{H2O}, CO, and \ce{N2} abundances outside their snowlines by a factor of a few compared to the initial abundances, and the efficient formation of \ce{CO2} ice, the abundance of which reaches more than 10 \% of the \ce{H2O} abundance within 1 Myr.
Note that \ce{CO2} ice is abundantly present in the ISM ice \citep[\ce{CO2}/\ce{H2O} $\sim$ 30 \%;][]{boogert15}, while there is no \ce{CO2} at the beginning of our simulations.
Therefore, the impact of the vertical mixing on the abundant ($>$10$^{-5}$ per H) ice molecules in the disk midplane is limited,
if ice in disks were inherited from the ISM ice.

\begin{figure*}[ht]
%    \centering
    \plotone{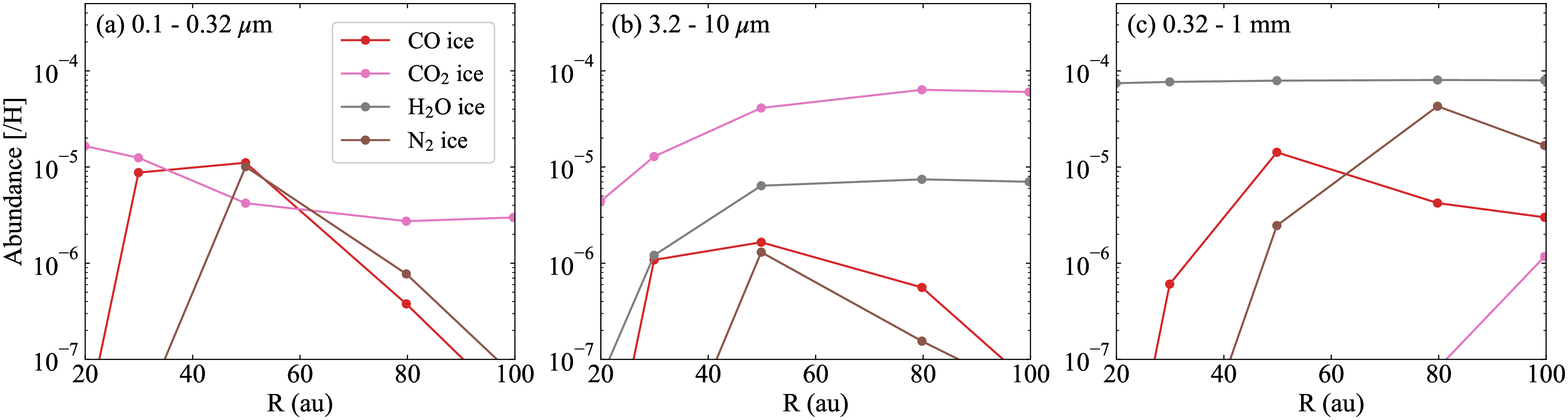}
    \caption{Radial profiles of abundances of selected icy molecules on small (0.1--0.32 $\mu$m), middle (3.2--10 $\mu$m), and large (0.32--1 mm) sized dust populations
    in the full model with $\alpha = 10^{-3}$ at 1 Myr.}
    \label{fig:iceab_mid_size}
\end{figure*}

On the other hand, if we look at the ice compositions of each dust population, the situation is different.
Figure \ref{fig:iceab_mid_size} shows radial profiles of abundances of selected icy molecules on small (0.1--0.32 $\mu$m), middle (3.2--10 $\mu$m), and large (0.32--1 mm) size dust populations in the full model with $\alpha = 10^{-3}$.
The ice compositions are very different, depending on dust size;
for the large dust population, the ice composition is dominated by \ce{H2O} ice as assumed in the initial conditions.
On the other hand, for the small- and middle-sized dust populations, the ice compositions are significantly different from the initial compositions; \ce{H2O} ice is no longer the most abundant icy species, and the \ce{CO2}/\ce{H2O} abundance ratio exceeds unity.
In reality, however, dust collision (coagulation and fragmentation) would average the ice mantle compositions and make the difference in the ice mantle compositions of different dust size smaller to some extent.
Coupled models with dust coagulation/fragmentation and gas-ice chemistry resolving the dust size distribution are necessary for better understanding the diversity of size-dependent ice compositions.

Laboratory experiments have shown that the surface composition of dust grains affect the threshold fragmentation velocity, and thus the efficiency of collisional growth.
In particular for this study, the threshold fragmentation velocity of \ce{CO2} ice is lower than that of \ce{H2O} ice \citep{musiolik16a,musiolik16b}.
The composition of the surface of ice mantles in disks would be determined by the thermal and chemical history of icy dust grains from its formation in molecular clouds to the delivery to disks \citep{drozdovskaya16,furuya17}, and the chemistry inside disks.
%Although our models do not consider the evolution in the earlier phases than the disk phase and assume arbitrary initial compositions of the disk, our models predict that the surface of ice mantles of small- and middle-sized dust are covered by \ce{CO2} ice.
Although our models do not track ice layered structure, judging from the efficient formation of \ce{CO2} ice, the surface of ice mantles of small- and middle-sized dust outside the CO snowline of disks may be covered by \ce{CO2} ice, regardless of the initial surface composition of the ice mantles in disks.
Therefore, the depletion of carbon and oxygen in the warm molecular layer may be also important for the dust coagulation/fragmentation in the disk midplane.

%Previous disk chemical models with grain surface chemistry and turbulent mixing \citep{semenov11,furuya13,furuya14} predicted more significant changes of the midplance ice compositions from the initial compositions than those predicted in this work.

\section{Summary} \label{sec:summary}
We investigated the temporal evolution of elemental abundances of volatile carbon, nitrogen, and oxygen in the warm molecular layer ($\gtrsim$20 K) of a protoplanetary disk, 
considering both the vertical cold finger effect and gas-ice chemistry.
We developed a one-dimensional model that incorporates dust settling, turbulent diffusion of dust and ices, as well as gas-ice chemistry including the chemistry driven by stellar UV/X-rays and the galactic cosmic rays.
Our main findings are summarized as follows.

\begin{enumerate}
{\item
Outside the CO snowline, the elemental abundance of carbon in the warm molecular layer can be reduced by a factor of $>$10 via the combination of the vertical cold finger effect of CO and the chemical conversion of CO into \ce{CO2} driven by stellar UV photons.
On the other hand, the gaseous \ce{N2}, the main nitrogen reservoir in the warm molecular layer, is less processed by the stellar UV driven chemistry, and exists as it is.
As the binding energy of \ce{N2} is lower than that of CO and \ce{CO2}, the degree of nitrogen depletion in the warm molecular layer is smaller than that of carbon and oxygen, 
leading to higher elemental abundance of nitrogen than those of carbon and oxygen.
}
{\item The degree of carbon and nitrogen depletion and the depletion timescale are independent of the assumed cosmic-ray ionization rate, at least, when $\xi_{\rm CR}$ is less than 10$^{-16}$ s$^{-1}$, but depends on the strength of turbulence.
In the case when the $\alpha$ value for the diffusion coefficient is 10$^{-3}$, 
the CO abundance in the warm molecular layers beyond the midplane CO snowline can become lower than the ISM abundance by a factors of $>$10 within 1 Myr.
The CO abundance continues to decrease until at 3 Myr.
On the other hand, the \ce{N2} abundance reaches steady state within 1 Myr.
When $\alpha = 10^{-2}$, both CO and \ce{N2} abundance reaches state state within 1 Myr.
When $\alpha = 10^{-4}$, the depletion factors of carbon and nitrogen are less than a factor of $\lesssim$5 in 1 Myr. 
}

{\item As a result of more significant depletion of elemental carbon and oxygen than nitrogen, the \ce{N2H+}/CO column density ratio increases with time in particular for the outer disk regions (outside the \ce{N2} snowline).
}

{\item 
In our models, elemental carbon-to-oxygen ratio in the gas phase ($\cogas$) becomes larger than 0.5, but does not exceed unity both inside and outside of the midplane CO snowline.
A promising mechanism for enhancing $\cogas$ would be the destruction of carbon grains by photoablation during the course of grain growth \cite[Section \ref{sec:column} and \ref{sec:compare}; see also][]{bosman21,vanclepper22}.
}

{\item The compositions of ice mantles at the midplane can be different, depending on dust size. Because the largest dust grains (1 mm in our models) are settled close to the midplane and the surface area per unit gas volume is small, their ice mantle compositions do not change significantly. On the other hand, smaller grains do not preserve their initial ice compositions.
}
{\item The size dependent dust temperature (i.e., larger grains have lower temperatures, assuming the radiative equilibrium) is crucial for the efficiency of the vertical cold finger effect (see the Appendix \ref{app:vertical_cold_finger} in details).
}

\end{enumerate}

\acknowledgments
We thank the anonymous referee for providing thoughtful and insightful comments that helped to improve the manuscript.
This work is supported in part by JSPS KAKENHI Grant numbers 18H05441, 19K03910, 20H00182, 20H05847 and 21K13967.
Numerical computations were in part carried out on PC cluster at Center for Computational Astrophysics, National Astronomical Observatory of Japan.
\software{RADMC-3D \citep{dullemond12}, \texttt{dsharp\_opac} \citep{birnstiel18}, Matplotlib \citep{matplotlib}}

\begin{appendix}
\renewcommand{\thefigure}{A\arabic{figure}}
\setcounter{figure}{0}
\renewcommand{\thetable}{A\arabic{table}}
\setcounter{table}{0}

\section{Importance of size dependent dust temperature for the vertical cold finger effect} \label{app:vertical_cold_finger}

%\epsscale{0.5}
\begin{figure*}[ht]
%    \centering
    \plotone{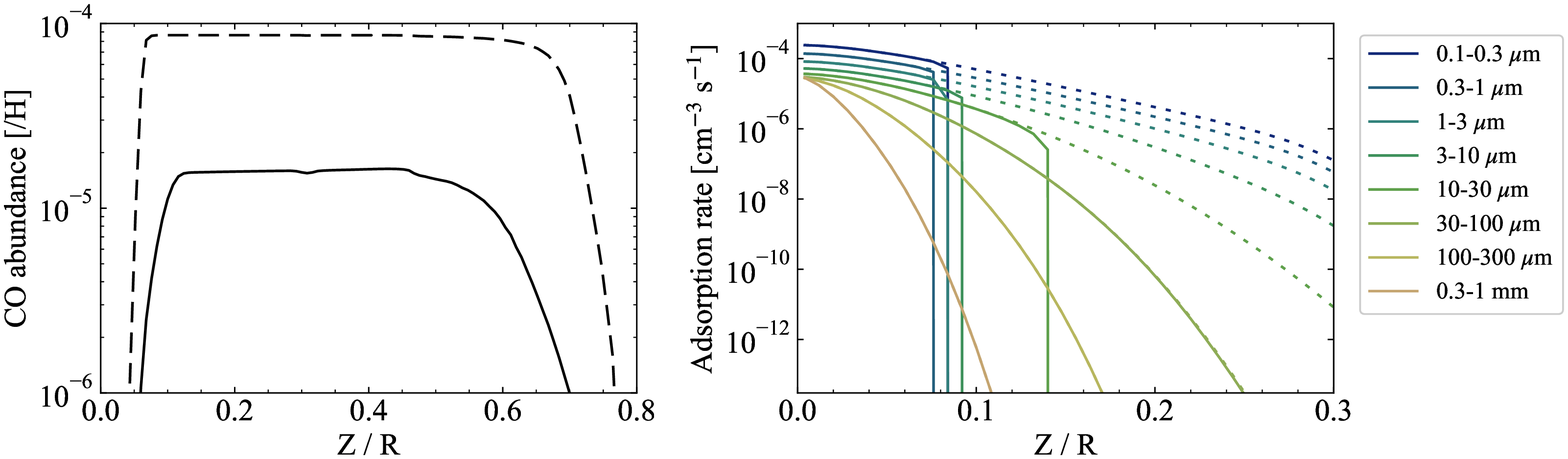}
    \caption{Left) Vertical distribution of the gas-phase CO abundance in the fiducial model (solid line) and in the single $T_d$ model (dashed line) at $R = 80$ au and at 1 Myr. In both models, grain surface chemistry is switched off.
    Right) Adsorption rates of CO onto 8 dust populations in the fiducial model are shown by dotted lines, while net adsorption rates (i.e., the adsorption rate of gas-phase CO minus the desorption rate of CO ice) are shown by solid lines.
    }
    \label{fig:co_vert}
\end{figure*}
%\epsscale{1.0}

In previous studies of the vertical cold finger effect of CO, the temperature of a dust grain is independent of its size, that is dust grains with different sizes have the same temperature \citep[e.g.,][]{kama16,xu17}.
Here, we show that this assumption can underestimate the impact of the vertical cold finger effect.

We ran the model without grain surface chemistry (see Section \ref{sec:result1}), employing two different assumptions for the dust temperature of the 8 dust populations.
In one model, dust temperature of dust populations are calculated in the same way as in our fiducial model.
Consequently, larger dust populations are colder than smaller ones.
In another model, all dust populations are assumed to have the same temperature (hereafter single $T_d$ model), and the dust temperature is assumed to be the area-weighted temperature averaged in the range between 0.1 $\mu$m and 1 mm (i.e., the entire size range in our fiducial models).
%Note that 1 mm-sized grain is the maximum size in the fiducial model, and thus has the lowest temperature.
The left panel of Figure \ref{fig:co_vert} compares the vertical distribution of the gas-phase CO abundance in the two models at $R=80$ au at 1 Myr.
The gas-phase CO abundance is much larger (i.e., the efficiency of the vertical cold finger effect is lower) in the single $T_d$ model compared to the fiducial model.
As discussed in \citet{kama16}, in the single $T_d$ model, the ratio of dust cross section contributed by large dust grains which are settled below the CO snow surface to the total dust cross section (denoted as $\Delta_X$) is the critical parameter for determining the efficiency of the vertical cold finger effect.
If $\Delta_X$ is small, most gas-phase CO adsorb onto small grains.
The CO-ice coated small grains can be transported above the CO snow surface by turbulent mixing, 
and the CO ice can sublimate into the gas phase. 
Thus the vertical cold finger effect is not efficient.
As shown in Fig. \ref{fig:phys}, the total cross section is dominated by small grains in our model.
Then the vertical cold finger effect is not efficient in the single $T_d$ model.

However, in the fiducial model, where smaller grains have higher temperatures, the net adsorption rate (i.e., the adsorption rate of gas-phase CO minus the desorption rate of CO ice) of the large dust grains can be larger than that of the small dust grains, even if $\Delta_X << 1$ (see the right panel of Fig. \ref{fig:co_vert}).
Then the single $T_d$ model underestimates the efficiency of the vertical cold finger effect.

The vertical cold finger effect was originally introduced by an analogy of the radial cold finger effect \citep{stevenson88,meijerink09}.
As the dust temperature is essentially the same in the disk midplane irrespective of its size, 
the size dependent dust temperature discussed here is not relevant for the radial cold finger effect, 
but is critically important for the vertical cold finger effect.

\section{Roles of different sized dust grains in disk chemistry} \label{append:ice}

\begin{figure*}[ht]
%    \centering
    \plotone{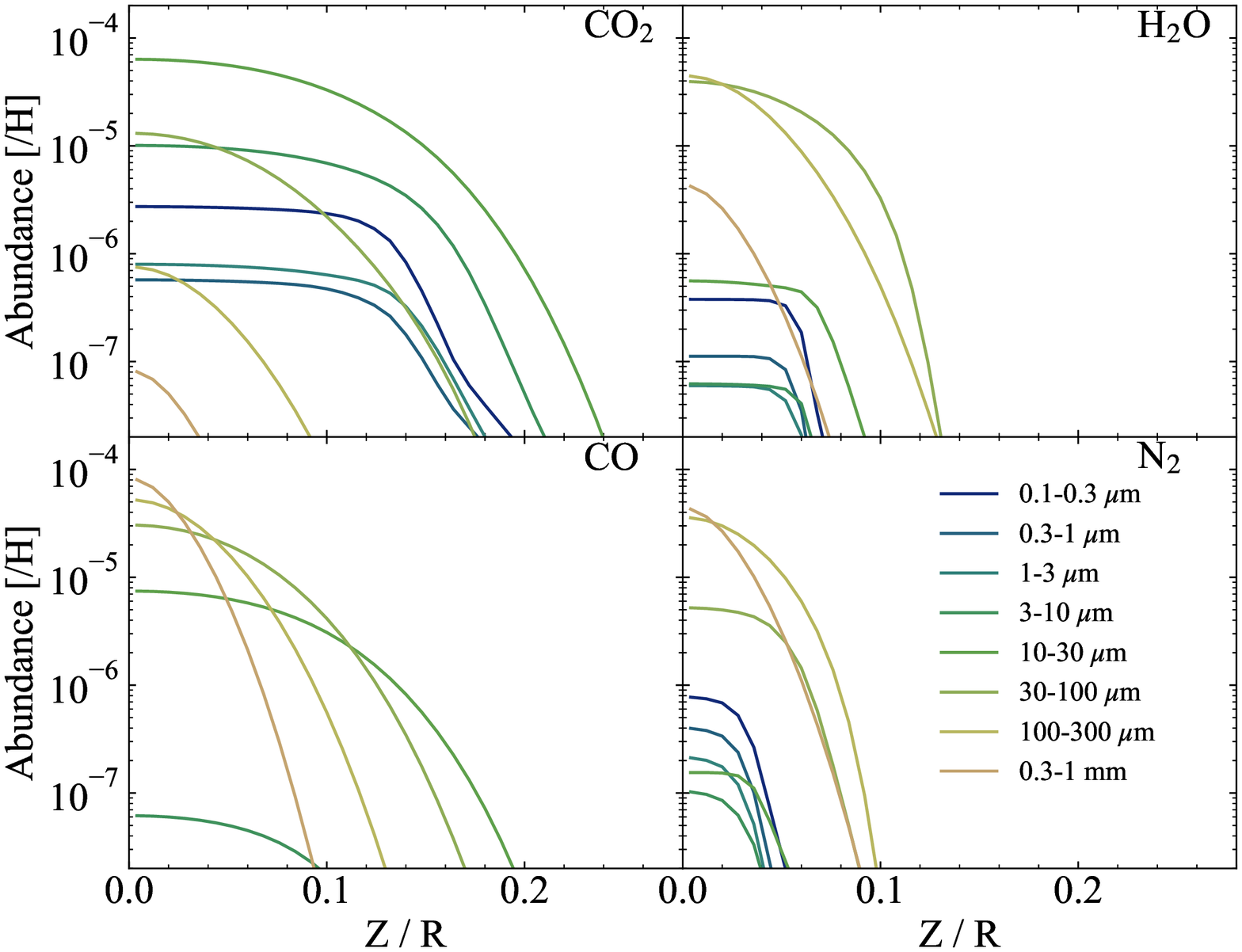}
    \caption{Abundances of \ce{CO2}, CO, \ce{H2O}, and \ce{N2} ices on 8 dust populations at $R=80$ au at 1 Myr in the full model.}
    \label{fig:80au_ice}
\end{figure*}

\begin{figure*}[ht]
%    \centering
    \plotone{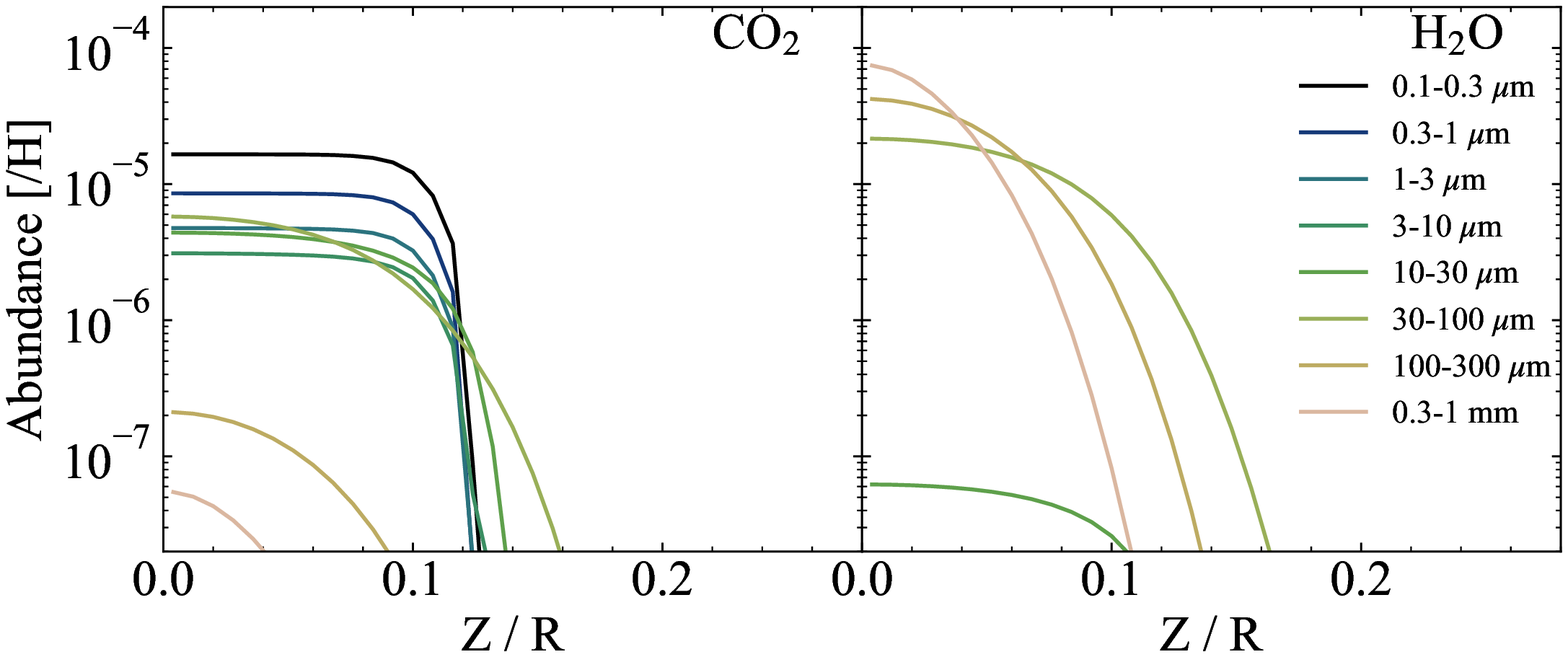}
    \caption{Similar to Figure \ref{fig:20au_ice}, but for $R=20$ au. As $R=20$ au is inside the CO snowline, only \ce{CO2} and \ce{H2O} abundances are shown.}
    \label{fig:20au_ice}
\end{figure*}

Figures \ref{fig:80au_ice} and \ref{fig:20au_ice} show the vertical distribution of the abundances of selected icy species on the 8 dust populations at $R = 80$ au and $R = 20$ au, respectively.
At $R = 80$, the formation of \ce{CO2} ice occurs mainly on the middle-sized grain populations ($\sim$10 $\mu$m), because gas-phase CO preferentially freeze out on those populations at $z/R \sim 0.1$ (right panel of Fig. \ref{fig:co_vert}), where stellar UV is not fully shielded.
Large dust populations ($\gtrsim$100 $\mu$m) are dynamically decoupled from the gas, 
and elements locked up in \ce{H2O}, CO, and \ce{N2} ices on the large dust grains do not contribute to  the active chemistry at $z/R \gtrsim 0.1$.

At $R=20$ au, on the other hand, the formation of \ce{CO2} ice occurs mainly on small dust populations ($\sim$1 $\mu$m or less), because the dust temperatures for all dust populations are too high for the CO freeze-out, and thus dust populations with larger total cross sections (i.e., small populations) are more important for the surface chemistry.
Then, the important dust size for \ce{CO2} formation depends on the distance from the star.
As at $R=80$ au, large dust populations ($\gtrsim$100 $\mu$m) are dynamically decoupled from the gas, and oxygen locked up in \ce{H2O} ice on the large dust grains do not contribute to the chemistry at $z/R \gtrsim 0.1$.

As discussed above, different sized dust grains contribute to the disk chemistry in the different ways.
Disk chemical models often adopt the single grain approximation, where gas-ice chemistry is described assuming a single-grain size and temperature \citep[see][and references therein]{gavino21}.
It is clear that the disk chemistry with turbulent mixing cannot be represented by models with the single grain approximation.
The similar argument was made by \citet{gavino21} for the case of static disk chemical models.

\section{Parameter dependence} \label{append:param}

\subsection{Cosmic-ray ionization rate}
In the models presented in Section \ref{sec:result}, the galactic cosmic-ray ionization rate was assumed to be $\xi_{\rm CR} = 10^{-18}$ s$^{-1}$.
Here we explore the impact of assumed values of $\xi_{\rm CR}$ on our results.
For this sake, we reran the static model and the full model with $\xi_{\rm CR} = 10^{-16}$ s$^{-1}$.
There has been no clear observational evidence of higher value of $\xi_{\rm CR}$ in disks than that in the dense ISM ($\sim$10$^{-17}$ s$^{-1}$).
\citet{cleeves15} concluded that the emission lines of \ce{HCO+} and \ce{N2H+} from the TW Hya disk are best reproduced by the model with low CR ionization rate of $10^{-19}$ s$^{-1}$.
\citet{seifert21} concluded that $\xi_{\rm CR}$ in the inner 100 au regions of the IM Lup disk is $10^{-20}$ s$^{-1}$ by modeling \ce{HCO+} and \ce{N2H+} emission, while that in the outer regions is $\gtrsim10^{-17}$ s$^{-1}$.
\citet{aikawa21} concluded that $\xi_{\rm CR} \gtrsim 10^{-18}$ s$^{-1}$ in the disks around IM Lup, AS 209, and HD 163296 on the basis of the comparisons between observationally derived column densities of \ce{HCO+}, \ce{N2H+}, and \ce{N2D+} and those predicted by a generic disk model \citep{aikawa18}.
Then the assumed value of $10^{-16}$ s$^{-1}$ may be considered as an upper limit of $\xi_{\rm CR}$ in the Class II disk environments.

\begin{figure*}[ht]
%    \centering
    \plotone{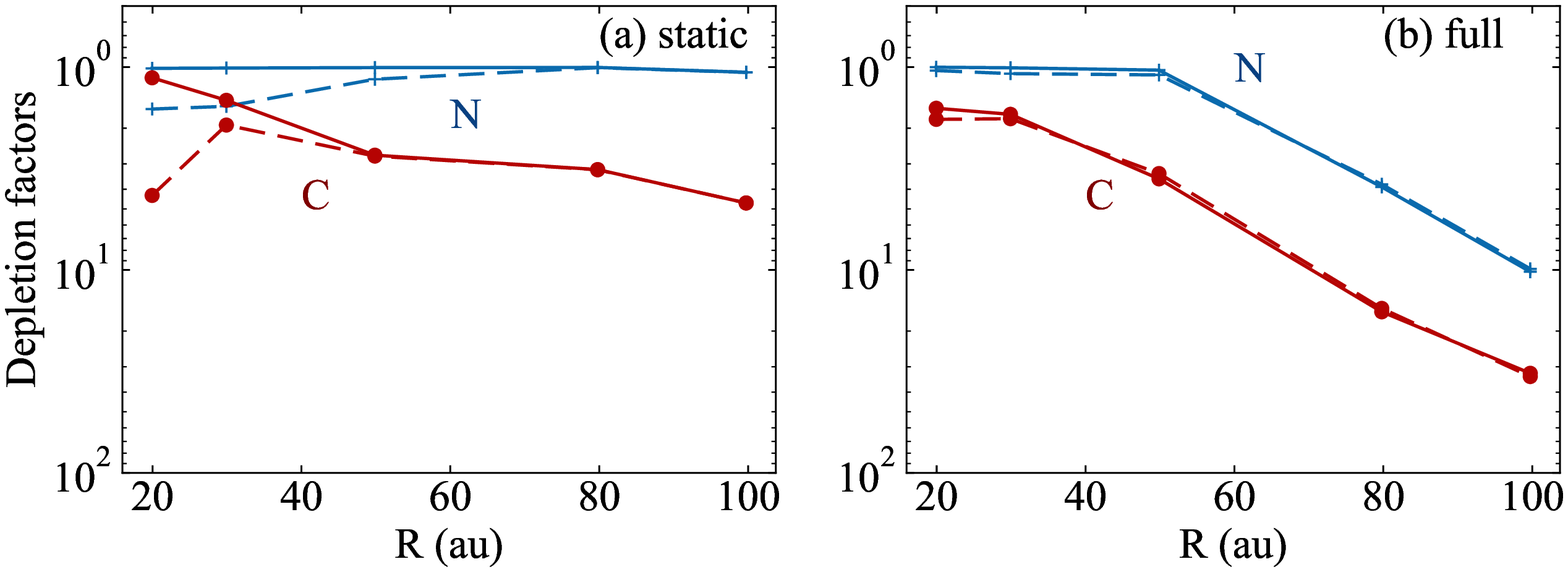}
    \caption{
    Depletion factors of elemental carbon (red lines) and nitrogen (blue lines) in the model with the enhanced CR ionization rate $\xi_{\rm CR}$ = 10$^{-16}$ s$^{-1}$ (solid lines) and the model with $\xi_{\rm CR}$ = 10$^{-18}$ s$^{-1}$ (dashed lines). 
    Left panel shows the static models, while right panel shows the full models.
    }
    \label{fig:dep_fac_append}
\end{figure*}

Figure \ref{fig:dep_fac_append} compares the depletion factors of carbon and nitrogen in the models with $\xi_{\rm CR} = 10^{-16}$ s$^{-1}$ with those in the models with $\xi_{\rm CR} = 10^{-18}$ s$^{-1}$ at 1 Myr.
In the static model case, the model with $\xi_{\rm CR} = 10^{-16}$ s$^{-1}$ shows higher depletion factors of carbon and nitrogen in particular inside the CO snowline ($\lesssim$30 au) and \ce{N2} snowline ($\lesssim$50 au), respectively, although the depletion factors are less than $\sim$10.
This is because the cosmic-ray driven chemistry converts gas-phase CO and gas-phase \ce{N2} near the midplane, where stellar UV and X-ray cannot penetrate, into less volatile icy molecules.
The timescale of the chemistry ($\tau_{\rm CRchem}$) is longer than 1 Myr when $\xi_{\rm CR} = 10^{-18}$ s$^{-1}$,
whereas the timescale is the order of 10$^5$ yr when $\xi_{\rm CR} = 10^{-16}$ s$^{-1}$ (see Section \ref{sec:timescale}).
In the full model case, the assumed value of $\xi_{\rm CR}$ does not significantly affect the depletion factors of carbon and nitrogen, because $\tau_{\rm CRchem}$ is larger than $\tau_{\rm phd}$ and $\tau_{\rm mix}$ even when $\xi_{\rm CR} = 10^{-16}$ s$^{-1}$.
Therefore, we conclude that the cosmic-ray ionization rate is less important for the chemistry of carbon (and nitrogen) depletion in the warm gas ($\gtrsim$20 K) than had been previously thought, when the dust grains have settled and UV radiation penetrate close to the midplane.
However, the cosmic-rays are the important parameter for determining the turbulence state of protoplanetary disks, since they control the ionization rate and the status of the magnetorotational instability \citep[e.g.,][]{sano00}.
Thus the cosmic-rays still indirectly play a key role in the carbon and nitrogen depletion through turbulence.

\subsection{Maximum and minimum sizes of dust grains}
In the models presented in Section \ref{sec:result}, the maximum size of dust grains ($a_{\rm max}$) is fixed to be 1 mm, as often assumed in disk chemical models \citep[e.g.,][]{zhang21}.
On the other hand, (sub)millimeter polarization observations of some disks have suggested that $a_{\rm max}$ is $\sim$100 $\mu$m \citep[e.g.,][]{kataoka16,kataoka17}.
Adopting the lower value of $a_{\rm max}$ affects the degrees of carbon and nitrogen depletion by several reasons.
First, the dust temperatures are lowered, and thus the efficiency of the vertical cold finger effect is increased.
Second, on the other hand, the relative fraction of the cross section of small dust grain, which are dynamically couple to the gas, to the total dust cross section increases, lowering the efficiency of the vertical cold finger effect.
Third, the mass fraction of large dust grains, which are dynamically decoupled from the gas, decreases, leading to larger amount of excess oxygen to convert CO to \ce{CO2} ice via the chemistry driven by stellar UV (Section \ref{sec:result1}).
Finally, the penetration of stellar UV photons is reduced, and the chemical conversion of CO into the \ce{CO2} ice becomes less efficient.

\begin{figure*}[ht]
%    \centering
    \plotone{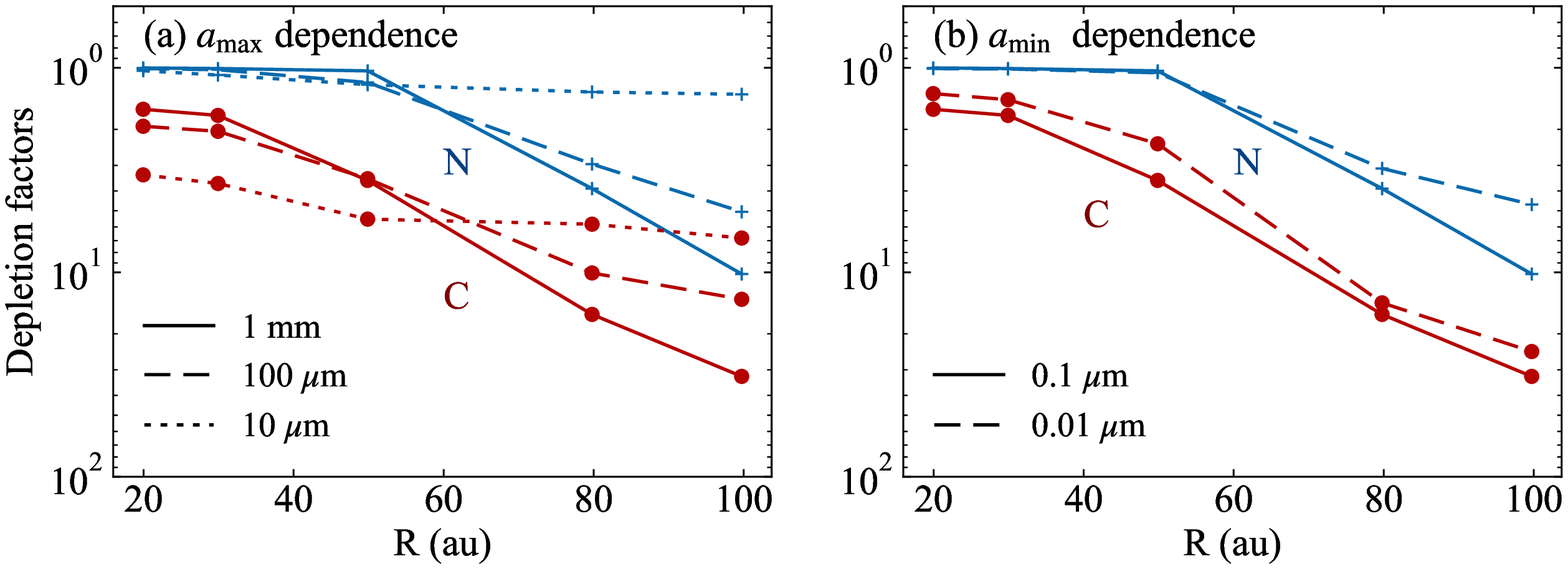}
    \caption{Dependence of depletion factors of elemental carbon (red lines) and nitrogen (blue lines) on $a_{\rm max}$ (left panel) and $a_{\rm min}$ (right panel) at 1 Myr.
    In the left panel, solid, dashed, and dotted lines represent the models with $a_{\rm max} = 1$ mm, 100 $\mu$m, and 10 $\mu$m, respectively.
    In the right panel, solid and dashed lines represent the models with $a_{\rm min}$ = 0.1 $\mu$m and 0.01 $\mu$m, respectively.}
    \label{fig:dep_fac_append2}
\end{figure*}

To check the impact of $a_{\rm max}$ on our results, we recalculated the entire modeling processes (Figure \ref{fig:flowchart}) with $a_{\rm max}$ = 100 $\mu$m or 10 $\mu$m as the value of $a_{\rm max}$ affects the disk physical conditions.
As shown in the left panel of Figure \ref{fig:dep_fac_append2}, the overall trends of the carbon and nitrogen depletion, that is the depletion of carbon is more significant than that of nitrogen, are not sensitive to the choice of $a_{\rm max}$, while the values for $\fdep{C}$ and $\fdep{N}$ are depend on $a_{\rm max}$.
%In particular, in the model with $a_{\rm max} = 10 \mu$m, $\fdep{C}$ and $\fdep{N}$ are much smaller than those in the model with $a_{\rm max}$ = 1mm, and $\fdep{C} \lesssim 5$ and $\fdep{N} \sim 1$.
At $R \gtrsim 50$ au, both $\fdep{C}$ and $\fdep{N}$ are lowered with reducing $a_{\rm max}$.
This indicates that the second effect is the most important for the outer regions ($\gtrsim50$ au), assisted by the forth effect in the case for carbon.
At $R \lesssim 50$ au, $\fdep{C}$ is enhanced with reducing $a_{\rm max}$, while $\fdep{N}$ is insensitive to $a_{\rm max}$.
This result indicates the third effect is the most important at $R \lesssim 50$ au.

The smallest dust grains (0.1 $\mu$m) in our models are not fully dynamically coupled with gas (see panel c in Figure \ref{fig:phys}).
To check the impact of the presence of even smaller dust grains, we also recalculated the entire modeling processes with $a_{\rm min}$ = 0.01 $\mu$m rather than $a_{\rm min}$ = 0.1 $\mu$m.
We confirmed that adopting the smaller value of $a_{\rm min}$ does not affect the depletion factors significantly (right panel of Figure \ref{fig:dep_fac_append2}).

%We find that the first effect, lower $a_{\rm max}$ leads to lower dust temperatures, is the most important one.
%In the model with lower $a_{\rm max}$, the position of the snowlines are closer to the central star compared to the model with larger $a_{\rm max}$. 
%In the model with $a_{\rm max}$ = 10 $\mu$m, the midplane snowlines of CO and \ce{N2} are located inside 20 au and at around 30 au, respectively, while they are located around 30 au in the model with $a_{\rm max}$ = 100 $\mu$m.
%In addition, the heights of the snow surfaces of CO and \ce{N2} are increased.
%Consequently, the vertical cold finger effect of CO and \ce{N2} are more efficient in the models with lower $a_{\rm max}$.

%\epsscale{0.5}
%\epsscale{1.0}

%\section{Effect of oxygen depletion due to dust settling}

\end{appendix}

\bibliography{ms}{}
\bibliographystyle{aasjournal}
%\begin{thebibliography}
%\end{thebibliography}

%% This command is needed to show the entire author+affilation list when
%% the collaboration and author truncation commands are used.  It has to
%% go at the end of the manuscript.
%\allauthors

%% Include this line if you are using the \added, \replaced, \deleted
%% commands to see a summary list of all changes at the end of the article.
%\listofchanges

%% For this sample we use BibTeX plus aasjournals.bst to generate the
%% the bibliography. The sample63.bib file was populated from ADS. To
%% get the citations to show in the compiled file do the following:
%%
%% pdflatex sample63.tex
%% bibtext sample63
%% pdflatex sample63.tex
%% pdflatex sample63.tex

%% This command is needed to show the entire author+affiliation list when
%% the collaboration and author truncation commands are used.  It has to
%% go at the end of the manuscript.
%\allauthors

%% Include this line if you are using the \added, \replaced, \deleted
%% commands to see a summary list of all changes at the end of the article.
%\listofchanges

\end{document}